# Clarifying $NH_2$ + $O(^3P)$ Reaction Dynamics: A Full-Dimensional MRCI, Machine-Learned PES Unravels High-Temperature Kinetics

Ying Xing, Weijie Hua, and Junxiang Zuo*

MIIT Key Laboratory of Semiconductor Microstructure and Quantum Sensing, Department of Applied Physics, School of Physics, Nanjing University of Science and Technology, Nanjing 210094, China

*Corresponding author: jxzuo@njust.edu.cn.

**Abstract**

The $NH_2$ + O reaction represents a critical oxidation pathway in ammonia and hydrazine combustion, yet significant discrepancies persist in reported kinetics. Here, we generate a full-dimensional ground-state potential energy surface (PES) for $NH_2O$ using high-level internally contracted multi-reference configuration interaction (ic-MRCI) calculations and the permutation invariant polynomial-neural network (PIP-NN) method. The PES encompasses all energetically accessible channels, including HNO + H, NH + OH, NO + $H_2$, and HON + H. Quasi-classical trajectory calculations on this surface yield thermal rate coefficients and branching ratios over a wide temperature range, particularly extending into the high-temperature regime relevant to combustion. The results provide accurate first-principles kinetic data essential for refining combustion models of nitrogen-containing fuels.

## I. Introduction

The prospective utilization of ammonia as a sustainable, carbon-free fuel has garnered significant recent interest, owing to its potential as a hydrogen carrier and its promise in decarbonizing energy systems.[1-5] The combustion chemistry of ammonia is fundamentally governed by the reactivity of key radical intermediates, among which the amino radical ($NH_2$) is paramount. In particular, kinetic sensitivity analyses have consistently identified the reaction of $NH_2$ with ground-state atomic oxygen, $O(^3P)$, as a highly sensitive pathway exerting considerable influence on key combustion properties, such as the laminar burning velocity.[6-10] A precise, quantitative understanding of the elementary reaction mechanisms in ammonia combustion is therefore not only of fundamental interest but is also crucial for optimizing combustion efficiency and advancing related technologies.

The $NH_2$ + $O(^3P)$ reaction proceeds through several energetically accessible product channels[11]:

$$\mathrm{NH_2(^2B_1) + O(^3P) \longrightarrow HNO(^1A'') + H,} \quad \text{(R1)}$$

$$\mathrm{\longrightarrow NH(^3\Sigma^-) + OH(^2\Pi),} \quad \text{(R2)}$$

$$\mathrm{\longrightarrow NO(^2\Pi) + H_2(^1\Sigma_g^+),} \quad \text{(R3)}$$

$$\mathrm{\longrightarrow HON(^3A'') + H.} \quad \text{(R4)}$$

Despite its critical importance, the kinetics and dynamics of this reaction are not yet fully resolved， especially at high temperatures (> 500 K). Early experimental studies, primarily conducted at low temperatures, established that the total rate coefficient is largely pressure-independent and that the HNO + H channel (R1) dominates, with a minor contribution from the NH + OH channel (R2).[12-15] At room temperature, Dransfield *et al.* [13] reported an overall rate coefficient of $(8.8\pm2.5)\times10^{-11}$ $cm^3\cdot molecule^{-1}\cdot s^{-1}$ using laser magnetic resonance and laser-induced fluorescence within a flow-tube reactor, which is about 25 times higher than that obtained in earlier flash photolysis study by Gehring *et al*[12]. Meanwhile, the product-specific rate coefficients for the R1 and R2 channels were quantified via computer modeling of radical concentration profiles as $(7.6\pm2.0)\times10^{-11}$ and $(1.2\pm0.5)\times10^{-11}$ $cm^3\cdot molecule^{-1}\cdot s^{-1}$, respectively.[13] Adamson *et al*.[14] proposed that the room-temperature rate coefficients for the $NH_2$ + O reaction is $(6.5\pm1.3)\times10^{-11}$ $cm^3\cdot molecule^{-1}\cdot s^{-1}$ and the branching into channel R2 is only 5-8% by infrared kinetic spectroscopy. Inomata and Washida[15] examined the temperature dependence of the reaction $NH_2$ + O and temperature-independent overall rate coefficients of $(1.2\pm0.3)\times10^{-10}$ $cm^3\cdot molecule^{-1}\cdot s^{-1}$ was observed over the temperature range 242-473 K in a few Torr. Crossed molecule beam scattering experiments reveal that both the OD and HNO products from $ND_2/NH_2$ + $O(^3P)$ reactions exhibit remarkably cold internal state distributions, indicating a significant exit barrier inconsistent with statistical predictions.[16, 17] Experimental data at elevated temperatures relevant to combustion is sorely lacking, and significant discrepancies persist among the limited existing studies.

Theoretical efforts to characterize the underlying potential energy surface (PES) of $NH_2O$ system have also been limited. Seminal work by Walch[18] employed complete active space self-consistent field (CASSCF) and internally contracted configuration interaction (ICCI) methods, while Yang *et al*.[19] and Wolf et al.[20] applied the Gaussian-2 (G2) methodology to map stationary points. Subsequent kinetic studies, such as those by Bozzelli and Dean[21] and Duan *et al.*[22], have relied on these earlier electronic structure data or simplified theoretical frameworks, inevitably constraining the accuracy of their predictions. Using the CCSD(T)/6-311++G(d,p)//UMP2/6-31G(d,p) level of theory, Sumathi *et al*.[23] identified a third reaction channel producing NO + $H_2$ (R3) and predicted equal branching ratios for R1, R2, and R3 at room temperature, a finding that conflicts with all existing experimental data. While the broader $NH_2O$ PES has been explored in the context of related reactions (e.g., H + HNO, NH + OH, and $H_2NO$ decomposition),[23-30] the insights from these studies are inherently limited by the computational methods available at the time, often failing to fully capture the multi-reference character and anharmonic effects critical for this system. In 2014, Homayoon and Bowman[31] reported the first global PES for reverse N($^2$D) + $H_2O$ reaction, constructed via a permutationally invariant linear least-squares fit to extensive UCCSD(T)-F12/aug-cc-pVTZ energies (supplemented by some MS-CASPT2 calculations in multireference regions), and used quasi-classical trajectory (QCT) calculations to analyze the internal state distribution of the NH + OH channel. Meanwhile, Isegawa *et al*.[32] carried out a CASPT2-based investigation into the reaction paths of N($^2$D) + $H_2O$ on the two lowest doublet states, specifically focusing on the key roaming pathways. Recently, Klippenstein *et al*.[11] investigated the $NH_2O$ PES using a high-level CCSD(T)-F12 and ANL1′ composite method, and derived the associated kinetics with variable reaction coordinate transition state theory (VRC-TST) alongside master equation simulations. Their findings demonstrate that, under combustion conditions, the $NH_2$ + O reaction primarily proceeds through HNO + H and NH + OH channels, with a previously overlooked NO + $H_2$ pathway contributing about 10% of the total reactive flux.

The development of accurate combustion models for nitrogen-containing fuels is fundamentally hindered by two interrelated limitations: the absence of a high-accuracy, full-dimensional PES for the pivotal $NH_2O$ system, and the consequent lack of rigorous dynamical investigations. Resolving this deficiency is essential for clarifying long-standing kinetic discrepancies and for generating reliable first-principles data under combustion-relevant conditions. To this end, we present a theoretical framework that seamlessly integrates state-of-the-art *ab initio* calculations, machine-learning driven PES construction, and high-throughput QCT simulations. Specifically, the electronic ground state of $NH_2O$ is characterized at the multi-reference configuration interaction (MRCI)[33, 34] level with large basis set. A global full-dimensional PES is then constructed by fitting over 62,000 *ab initio* energy points using the permutation-invariant polynomial-neural network (PIP-NN)[35, 36] approach. This novel PES provides the robust foundation for exhaustive QCT calculations to elucidate the reaction mechanism, quantify accurate thermal rate coefficients, and determine branching ratios across a wide temperature range. This comprehensive methodology enables us to offer unprecedented insight into the microscopic reaction dynamics of $NH_2$ + O and provide foundational kinetic data essential for refining combustion models of nitrogen-containing fuels.

The remainder of this paper is organized as follows: Section II details the computational methodology, including the *ab initio* calculations, PES construction, and dynamics simulations. The results and a thorough discussion are presented in Section III, and the principal conclusions are summarized in Section IV.

## II. Computational Methods

### A. *Ab initio* calculations

This study focuses on the ground doublet electronic state ($^2A'$) of the $NH_2O$ system. $NH_2$ + O reaction constitutes a prototypical radical-radical encounter, for which a quantitative assessment of multireference character across different regions of the PES is essential. To this end, we employed the M diagnostic[37], which is based on natural orbital occupation numbers, to evaluate the extent of strong electronic correlation. The computed M values for all stationary points on the $NH_2O(^2A')$ PES are summarized in Table I. These results reveal pronounced multireference character (M > 0.10) at every stationary point, confirming the necessity of a multireference treatment for this reaction system.

Accordingly, all electronic structure calculations were performed using the dynamically weighted state-averaged complete active space self-consistent field (DW-SA-CASSCF) method[38, 39]. Specifically, the active space comprised 9 electrons in 8 orbitals, and three-state ($3\ ^2A'$) DW-SA-CASSCF calculations were carried out with the weighting function $w(i) = 1/\cosh^2(\beta\Delta E)$, where $\Delta E = E_i - E_0$ is the energy difference between the $i$-th state and the ground state, and $\beta^{-1} = 3.0$ eV. The resulting reference wave functions were subsequently employed in explicitly correlated multi-reference configuration interaction (MRCI-F12)[34] calculations, in which higher-order excitation effects and size-extensivity corrections were accounted for via the rotated-reference Davidson (+Q) scheme[40]. To ensure compatibility with the explicitly correlated formalism and to balance computational cost with accuracy, the correlation-consistent polarized valence triple-zeta basis set designed for F12 methods (cc-pVTZ-F12)[41] was employed throughout. For comparative validation, the reactants and four products structures were also optimized at the unrestricted coupled-cluster level with single, double, and perturbative triple excitations using the explicitly correlated F12a approximation (UCCSD(T)-F12a)[42, 43]. All single-point energy calculations required for constructing the global PES were carried out at the MRCI-F12+Q/cc-pVTZ-F12 level of theory using the MOLPRO program package[44].

**B. Construction of the potential energy surface**

The global full-dimensional PES was constructed using the PIP-NN approach, which provides highly accurate representation of PESs with build-in permutation invariance. The fitting database was generated through an initial sampling of relevant configuration space via direct dynamics calculations performed at the UB3LYP/6-311+G(d,p) level. From this initial dataset, representative geometries were selected, and their single-point energies were computed at the high-level MRCI-F12+Q/cc-pVTZ-F12 theory. A preliminary PES was first fitted using the PIP-NN method. Subsequently, an ensemble of quasi-classical trajectories was propagated on this primitive PES to dynamically sample regions critical to the reaction dynamics with initial conditions sampled from thermal distributions of vibrational, rotational, and translational energies. Candidate geometries were selected based on two criteria: (i) energies below 5 eV relative to the $NH_2$ + O($^3$P) asymptotic limit, and (ii) significant deviation among the three best-fitted PESs. To avoid excessive point density across the configuration space, a new geometry was retained only if its Euclidean distance[45] (in the space of internuclear distances) from all existing points exceeded 0.15 Å. This iterative refinement process resulted in a final database of approximately 62,000 geometries, which was used to obtained the converged global PES.

The permutation symmetry of the two hydrogen atoms was encoded in the PES through a set of 49 permutationally invariant polynomials (PIPs)[46] up to third-order, which served as the input to the neural network (NN). The PIPs were built from Morse-like variables of the form $p_{ij} = \exp\left(-r_{ij}/\alpha\right)$, where $r_{ij}$ denotes the internuclear distance between atoms $i$ and $j$, and $\alpha$ was set to 1.5 Å. A standard feedforward NN with the architecture 49–30–50–1 (input layer, two hidden layers, output layer) was adopted. The hidden layers employed hyperbolic tangent activation functions, while a linear activation was used in the output layer. The dataset was randomly partitioned into training (90%), validation (5%), and testing (5%) subsets. Network training was conducted using the Levenberg–Marquardt algorithm[47] in conjunction with "early stopping" criterion[48] to minimize the root-mean-square error (RMSE), defined as $\mathrm{RMSE} = \sqrt{\sum_{i=1}^{N_{\mathrm{data}}}\left(E_i^{\mathrm{PES}} - E_i^{ab\ initio}\right)^2 / N_{\mathrm{data}}}$, where $E_i^{\mathrm{PES}}$ and $E_i^{ab\ initio}$ denote the fitted and *ab initio* energies of the *i*-th data point, respectively. This strategy effectively prevents overfitting during the training process. A total of 100 independent neural networks were trained and the final PES was constructed by averaging the predictions of the three best-performing networks in the ensemble, thereby enhancing both robustness and predictive reliability.

**C. Quasi-Classical Trajectory Calculations**

QCT calculations for the $NH_2$ + O reaction were carried out on the newly developed global PES using the VENUS96 program package.[49] Trajectories were initiated with a reactant separation of 10.0 Å and terminated when any product fragment reached 15.0 Å or when non-reactive trajectories results in a reactant separation of 15.0 Å. The maximum impact parameter, $b_{\mathrm{max}}$, was determined as 8.5 Å based on 10,000 preliminary trajectories at 200 K. A time step of 0.1 fs was used for numerical integration. The total energy was conserved within 0.01 kcal/mol for the vast majority of trajectories across all temperatures, with the proportion of trajectories meeting this criterion ranging from 95.06% at 200 K to 98.78% at 2500 K. This high level of energy conservation confirms the numerical smoothness and stability of the PES during dynamical integration. To mitigate the inherent zero-point energy (ZPE) leakage issue in QCT simulations, a symmetric "passive" ZPE correction scheme[50] was applied, in which trajectories ending with any reactants or products vibrational mode fell below its corresponding ZPE were excluded from the reactive event count.

The thermal rate coefficient, $k(T)$, was determined as:

$$k(T) = g_{\mathrm{e}} \left( \frac{8 k_{\mathrm{B}} T}{\pi \mu} \right)^{1/2} \pi b_{\mathrm{max}}^2 \frac{N_{\mathrm{r}}}{N_{\mathrm{tot}}}. \quad (1)$$

Here, $\mu = (m_{\mathrm{NH_2}} m_{\mathrm{O}}) / (m_{\mathrm{NH_2}} + m_{\mathrm{O}})$ is the reduced mass of the reactants, $k_{\mathrm{B}}$ is the Boltzmann coefficient, and $N_{\mathrm{r}}$ and $N_{\mathrm{tot}}$ denote the numbers of reactive and total trajectories, respectively. An electronic degeneracy factor[51] was incorporated to account for the electronic partitions of the reactants and the adiabatic PES:

$$g_e = \frac{q_{NH_2O}}{q_{NH_2} q_O}. \quad (2)$$

The electronic partition functions are: $q_{NH_2O} = 2$ for the $^2A'$ state; $q_{NH_2} = 2$ for the $NH_2(^2B_1)$; and $q_O = 5 + 3e^{-158.5cm^{-1}/k_BT} + e^{-226.5cm^{-1}/k_BT}$ for the spin–orbit split triplet $O(^3P)$. The statistical uncertainty is evaluated as $\Delta = [(N_{tot} - N_r)/(N_{tot}N_r)]^{1/2}$.

## III. Results and Discussion

### A. Multireference Diagnostics

The optimized geometries of asymptotes and stationary points, obtained at the MRCI-F12/cc-pVTZ-F12 level, are provided in Table II, with full Cartesian coordinates available in the Supporting Information. Their corresponding relative energies and harmonic vibrational frequencies are summarized in Table III, along with values from our fitted PES, the UCCSD(T)-F12a/cc-pVTZ-F12 calculations, and available experimental benchmarks[52]. Following ZPE correction, the MRCI-F12-derived reaction energies for the four product channels exhibit significantly better agreement with experiments than the UCCSD(T)-F12a results. The respective deviations from experiments for the HNO + H, NH + OH, NO + $H_2$ and HON + H channels are 0.13, 0.17, 0, and 0.29 kcal/mol at the MRCI-F12 level, versus 2.71, 0.25, 2.91, and 1.55 kcal/mol at the UCCSD(T)-F12a level.

For comprehensive comparison, Figure 1 systematically illustrates the deviations in stationary point energies reported in prior theoretical studies relative to the present MRCI-F12 benchmark. A striking observation is the pronounced sensitivity of the predicted energetics to the choice of electronic structure method, particularly for the deeply bound intermediates and several transition states. Deviations can reach nearly 10 kcal/mol, underscoring the inherent difficulty in achieving quantitative accuracy for this open-shell multi-radical system. When grouped by theoretical approach, the deviations exhibit discernible trends. Studies employing single-reference methods, such as G2 theory[19] and CCSD(T)/CBS[53], consistently predict the $H_2NO$ intermediate to be less stable (higher in energy) than our MRCI-F12 result. This is likely attributable to the inadequate description of static correlation in this region, where the multireference character is significant (M ≈ 0.39, Table I). Conversely, methods with explicit multireference treatment, such as the CASPT2 calculations of Isegawa *et al*.[32], show substantially better agreement for key isomers like *trans*-HNOH and *cis*-HNOH. Nevertheless, even among multireference methods, variations in active space selection and dynamic correlation treatment lead to non-negligible discrepancies, as seen in the earlier MRCI+QC study by Walch[18]. The most substantial deviations occur for the transition states TS4 and TS5, which govern the critical HNO + H formation channels. The generally robust CCSD-F12 calculations[31] slightly overestimates the barrier for TS4 by 2.47 kcal/mol, while the ANL1′ composite method[11] underestimates it by 5.54 kcal/mol. We attribute this to these differences to the challenging electronic structure at these geometries, where bond-breaking and forming processes involve significant diradical character, as reflected by the high M diagnostics (~0.58–0.60).

In addition to the M diagnostic, we have computed a set of complementary multireference diagnostics ($T_1$, $D_1$, and $D_2$), with full results available in Table S1 of the Supporting Information. A consistent trend is observed across these independent metrics and structures with pronounced multireference character (M > 0.10) typically exhibit $T_1$ values exceeding 0.02 and significant $D_1/D_2$ weights. For instance, regions with high multireference indices M (e.g., TS4/TS5) are consistently identified by elevated $T_1$ amplitudes (> 0.05) and notable contributions from secondary determinants ($D_1$ > 0.14, $D_2$ > 0.21). These diagnostics provide robust evidence that a multireference treatment is indispensable for achieving quantitative accuracy throughout the entire reaction coordinate. Single-reference coupled-cluster methods, unless explicitly corrected for multireference effects, are known to perform poorly in such regions, which may potentially lead to skewed energetic rankings and thus unreliable branching ratio predictions. The excellent agreement between our MRCI-F12+Q energetics and available experimental reaction enthalpies (Table III) further validates the current methodological strategy. By employing a DW-SA-CASSCF reference that captures near-degeneracy effects, followed by an explicitly correlated MRCI treatment, we achieve a balanced and accurate description across the full configuration space.

## B. Features of the Potential Energy Surface

A total of 62,452 *ab initio* points calculated at the MRCI-F12 level were employed to train the global PIP-NN PES for the multichannel $NH_2$ + O reaction. This sampling profile results directly from our iterative, dynamics-guided sampling strategy. As shown in Figure 2(a), the energy distribution of these points is markedly unimodal, with approximately 70% lying within ±20 kcal/mol of the reactant asymptote. Additionally, the distribution extends broadly from the reactant asymptote down to the deep $H_2NO$ minimum (~ -90 kcal/mol), while a higher-energy wing reaches up to +5 eV (~ 115 kcal/mol) relative to reactants. This comprehensive coverage ensures that the PES faithfully captures not only minima and transition states but also repulsive walls and non-reactive encounters essential for computing accurate cross sections and thermal rate coefficients.

The fidelity of the fitted PIP-NN PES quantitatively assessed in Figures 2(b) and 2(c). The overall RMSE is 0.63 kcal/mol and the maximum absolute deviation is 20.21 kcal/mol. Crucially, as shown in Figure 2(b), the vast majority of fitting errors lie within a narrow symmetric band of ±1 kcal/mol, which is well within the accepted threshold of “chemical accuracy” for high-level dynamics simulations. Figure 2(c) further reveals that the error distribution is sharply peaked near zero, with 62.1% of points having errors below 0.1 kcal/mol and over 97% of points below 1 kcal/mol. This indicates that the PIP-NN functional form combined with permutationally invariant input is highly effective in capturing the underlying quantum chemical data. The observed maximum deviations (up to 20 kcal/mol) are confined to energetically high-lying, kinetically inaccessible regions of configuration space and do not affect the computed reaction probabilities or rate coefficients under the thermal and geometric constraints of our simulations.

Figure 3 presents the schematic energy diagram for the $NH_2$ + O reaction on the ground-state ($^2A'$) PES, with energies obtained from both the fitted PES (red) and direct *ab initio* calculations (blue). The excellent agreement between the two sets of values across all stationary points confirms the high fidelity of our PIP-NN PES in capturing the intricate energetics of the system. The diagram reveals a complex but well-defined reaction network centered on three stable minima: the initial association complex $H_2NO$, and the two isomers *trans*-HNOH and *cis*-HNOH. These species are interconnected via isomerization transitions states (TS1, TS2) and serve as gateways to the four product channels (HNO + H, NH + OH, NO + $H_2$, and HON + H) through distinct dissociation transition states (TS3–TS5).

The reaction initiates via the barrierless association of $NH_2$ and O, leading to the formation of a deeply bound $H_2NO$ intermediate located 90.28 kcal/mol below the reactant asymptote. The pronounced depth of this well implies highly efficient initial capture and substantial intramolecular energy redistribution, suggesting that statistical theories may have some validity at the early stage. The subsequent fate of the $H_2NO$ complex is governed by competing isomerization and dissociation pathways with substantial barriers. The isomerization to *trans*-HNOH via a 1,2-hydrogen shift through TS1 requires overcoming a barrier of 53.29 kcal/mol relative to $H_2NO$. This pathway remains a feasible process under combustion temperatures, though not instantaneous.

Alternatively, $H_2NO$ may undergo direct dissociation along two competing routes. One route leads to the highly exothermic NO + $H_2$ channel (-82.63 kcal/mol) through a synchronous or nearly synchronous double hydrogen migration over TS3, with a high barrier of 66.73 kcal/mol. The other forms HNO + H via TS5, with a comparable barrier of 67.67 kcal/mol. Despite their similar barrier heights, TS3 leads to a much more stable product set.

The isomerization between *trans*- and *cis*-HNOH proceeds through TS2, which lies only 13.79 kcal/mol above the *trans* conformer. Both HNOH isomers can further dissociate to several common product channels. These include direct cleavage to NH + OH or HON + H, as well as hydrogen elimination via TS4 to yield HNO + H. The barrier via TS4 is comparable to that of TS5, suggesting that HNO + H products can originate either from the HNOH isomerization pathway or from direct dissociation of $H_2NO$. The NH + OH channel proceeds through essentially barrierless N–O bond cleavage from the intermediate minima. Its yield is thus governed mainly by the competition between dissociation and isomerization out of the HNOH wells, as well as by the initial branching between $H_2NO$ and the TS1 pathway. Notably, all five transition states lie energetically below the separated $NH_2$ + O asymptote. Furthermore, all product channels are exothermic except for HON + H, which is slightly endothermic by 0.16 kcal/mol. While accessible from HNOH without a barrier, the HON + H channel is both kinetically and thermodynamically disfavored.

The accuracy of the fitted PIP-NN PES is further scrutinized beyond stationary points by examining the minimum energy paths (MEPs) for the key elementary steps governing the reaction network. Figure 4 presents a direct comparison between the MEPs traced on the PIP-NN PES and the benchmark *ab initio* single-point energies along the intrinsic reaction coordinate (IRC) for six critical processes. The near-perfect alignment observed across all panels [(a) ~ (f)] unequivocally demonstrates that our machine-learned PES possesses high fidelity in dynamic interpolation.

Figure 5 provides a two-dimensional "entrance channel" view of the $NH_2$ + O reaction, mapping the potential energy as the O atom approaches the rigid $NH_2$ radical in its equilibrium geometry. With the $NH_2$ plane fixed and the O atom scanned across it, the resulting contour plot highlights the pronounced anisotropy and barrierless nature of the initial interaction. A prominent feature is a broad deep attractive basin (blue contours, reaching approximately -80 kcal/mol), which directly illustrates the driving force for the barrierless association that forms the deep $H_2NO$ intermediate. The PES faithfully captures the high-level *ab initio* data across the configuration space critical to the reaction dynamics.

### C. Thermal Rate Coefficients and Branching Ratios

Based on the newly constructed $NH_2O(^2A')$ PES, QCT calculations were performed to investigate the kinetics of the $NH_2$ + O reaction at temperatures of 200, 298, 500, 700, 1000, 1500, 2000, and 2500 K. Approximately $10^5$-$10^6$ trajectories were run at each temperature, with statistical uncertainties kept below 0.47%. The total energy was conserved within $10^{-4}$ kcal/mol for the vast majority of trajectories, confirming the numerical smoothness and stability of the PES during dynamical integration.

The total thermal rate coefficients obtained from QCT simulations over the temperature range 200-2500 K are summarized in Table IV and displayed in Figure 6, alongside available experimental data[13-15] and earlier theoretical predictions[11, 23]. While experimental measurements are limited and predominantly lie between 200 and 500 K, the present rate coefficients calculated on the new PIP-NN PES exhibit a monotonic decrease with increasing temperature, spanning nearly an order of magnitude from $2.02\times10^{-10}$ $cm^3\cdot molecule^{-1}\cdot s^{-1}$ at 200 K to $3.31\times10^{-11}$ $cm^3 \cdot molecule^{-1} \cdot s^{-1}$ at 2500 K. At 298 K, our results are approximately 2-4 times larger than the scattered earlier measurements by Dransfield *et al.*[13] and Adamson *et al.*[14]. Importantly, within the temperature window covered by experiment (200-500 K), our computed rate coefficients of $1.63\times10^{-10}$ $cm^3 \cdot molecule^{-1} \cdot s^{-1}$ at 298 K and $1.04\times10^{-10}$ $cm^3 \cdot molecule^{-1} \cdot s^{-1}$ at 500 K are in excellent agreement with the most comprehensive experimental study by Inomata and Washida[15], who reported a temperature-independent value of $(1.2\pm0.3)\times10^{-10}$ $cm^3 \cdot molecule^{-1} \cdot s^{-1}$. This close agreement under well-characterized conditions validates the current theoretical framework and justifies its application for predicting kinetics at higher combustion-relevant temperatures, where experimental data remain scarce.

Meanwhile, the two prior theoretical studies show markedly different temperature dependence of the rate coefficient. Our results, which display a monotonic decrease, help to resolve this notable discrepancy. The work by Sumathi *et al.*[23], based on lower-level CCSD(T)//UMP2 calculations and quantum Rice-Ramsperger-Kassel theory (QRRK) analysis, predicts a non-physical "turnover" behavior in rate coefficient, characterized by an increase below 500 K followed by a decrease at higher temperatures. We attribute this artifact to inaccuracies in their underlying PES topology, specifically in the relative energies of key transition state, which subsequently skewed the statistical weights of competing reaction channels. In contrast, the recent high-level study by Klippenstein *et al.*[11], using a ANL1′//CCSD(T)-F12 composite PES within a VRC-TST/master equation framework, predicts a shallow minimum near 800 K, with rate coefficient rising toward both lower and higher temperatures. While the increase at high temperatures is modest, the qualitative divergence from our monotonically decreasing trend remains significant. We propose that this difference stems primarily from the quantitative description of the initial barrierless association step. Our QCT simulations, propagated on a full-dimensional PES, inherently capture the complete anisotropy and long-range interactions of the entrance channel (Figure 5), yielding a capture rate that decreases classically with rising temperature. Their sophisticated VRC-TST approach, however, may have different sensitivities to the parameterized long-range potentials and the treatment of angular momentum conservation.

Beyond 500 K, our work provides the first full-dimensional QCT predictions for thermal rate coefficients under combustion-relevant conditions. The monotonic decrease observed up to 2500 K indicates that the association-dominated mechanism remains operative across this entire range. The convergence of rate coefficients to a near-plateau value above 2000 K may indicate the approach to a gas-kinetic collision limit under these conditions. Although direct experimental validation at high temperature is currently lacking, the physical consistency of the calculated trend and the demonstrated accuracy of our PES at stationary points lend credence to these predictions. We therefore strongly recommend that the rate expression for $NH_2$+ O → products in detailed kinetic models of ammonia combustion be updated based on our computed values (Table IV) to capture a pronounced negative temperature dependence, especially above 1000 K.

The temperature-dependent thermal rate coefficients for the four individual product channels of the $NH_2$ + O reaction, derived from our QCT simulations, are provided in Table IV and compared with earlier experimental[13-15] and theoretical[11, 21-23, 54, 55] results in Figure 7. The rate for the HNO + H channel decreases monotonically with temperature, consistent with the overall total rate. This behavior arises because its major formation pathway proceeds through the long-lived $H_2NO$ intermediate, whose initial barrierless association is the rate-limiting step. The temperature-dependent decrease in capture efficiency thus dictates the observed trend for this channel. Notably, at low temperature (< 500 K) [Figure 7(a)], our calculated rate coefficients for the HNO + H channel aligns remarkably well with the experimental data reported by Inomata and Washida[15], which represent the most definitive product-specific measurements available in this regime. This agreement validates not only the total rate but also the fidelity of our PES and dynamics treatment in accurately partitioning the reactive flux among competing pathways.

As shown in Figures 7(b) and 7(c), the NH + OH and NO + $H_2$ channels both exhibit a decreasing trend with temperature, albeit through distinct underlying mechanisms. This branching behavior is a direct consequence of the potential energy landscape. The NH + OH channel proceeds via nearly barrierless N–O bond cleavage from the HNOH isomers. The fact that its absolute rate also declines with temperature indicates that the branching into the HNOH isomer pool via TS1 from $H_2NO$ does not increase sufficiently to compensate for the overall decrease in capture rate. Dynamically, at higher temperatures a larger fraction of $H_2NO$ complexes may dissociate directly (e.g., via TS5) or undergo other competing processes before isomerizing, thereby suppressing the growth of the NH + OH flux. In contrast, the NO + $H_2$ channel requires passage through the geometrically constrained tight TS3 directly from $H_2NO$. The declining population of the $H_2NO$ complex with rising temperature, combined with a presumably weak positive temperature dependence for crossing TS3, results in an overall decrease in this channel. Our predicted monotonically decreasing rates for both NH + OH and NO + $H_2$ contrast with several prior theoretical studies[11, 21-23, 54, 55], which reported more complex or even increasing trends. This discrepancy likely originates from our full-dimensional, explicitly dynamics QCT approach, which naturally captures the competition among all accessible pathways without relying on pre-assumed statistical weights. Critically, all product-specific rates decrease with temperature, implying that the overall consumption of $NH_2$ radical via reaction with O-atom becomes less efficient at high temperature, a crucial fact for predicting flame speeds and $NO_x$ yields in ammonia combustion.

The HON + H channel exhibits a unique, shallow minimum near 700 K [figure 7(d)]. At very low temperatures, it may benefit marginally from longer complex lifetimes. At high temperatures, its slight rise likely stems from an increased probability of accessing this high-energy, direct dissociation pathway from vibrationally excited HNOH isomers once the available energy exceeds the reaction endothermicity. This subtle non-monotonic behavior is a fine detail that only high-throughput, accurate dynamics on a global PES can capture, underscoring the sophistication of the present methodology. We thus report the first set of channel-resolved rate coefficients that consistently reflect this physically coherent picture.

The corresponding product branching ratios across the studied temperature range (200–2500 K) are summarized in Table V. Across the entire temperature range, the HNO + H channel remains dominant, contributing 47–70% of the total yield. The NH + OH channel represents a significant secondary pathway (15-39%), while the NO + $H_2$ channel provides a minor but non-negligible contribution (10-15%). The HON + H channel remains negligible throughout, never exceeding 3% of the total product yield. Overall, the branching ratios exhibit only moderate temperature dependence. Specifically, the HNO + H and NO + $H_2$ factions gradually decline from 70.30% and 14.36% at 200 K to 47.43% and 10.69% at 2500 K, respectively, whereas the NH + OH and HON + H factions increase from 15.00% and 0.59% to 38.97% and 2.83% over the same range. The growth in NH + OH branching fraction with temperature does not arise from an increase in its absolute rate, but rather from its slower decline relative to the dominant HNO + H channel. At low temperatures, the reaction is strongly dominated by the HNO + H channel, with NO + $H_2$ and NH + OH channels sharing most of the remaining yield. At elevated temperatures, although HNO + H remains the primary channel, the NH + OH pathway increases substantially and approaches it in overall contribution.

In the present QCT simulations, a symmetric passive ZPE leakage correction scheme (denote as SZPE) was implemented by discarding all trajectories violating the ZPE criterion. This approach ensures statistical consistency by applying the same physical constraint to both the numerator and denominator in the rate expression. Recognizing that ZPE handling remains a subject of ongoing discussion in QCT simulations, we performed two additional sets of calculations to assess the sensitivity of our results to different ZPE treatments: (i) no ZPE correction (NZPE), and (ii) a refined correction (RZPE) in which only reactive trajectories with ZPE-violating products were rejected while all non-reactive are retained regardless of the internal energy of the $NH_2$ fragment. The resulting thermal rate coefficients and product branching ratios for these two cases are compiled in Tables S2-S5 of the Supporting Information, and direct comparisons among the three treatments are presented in Figures S1-S3.

As shown in Figure S1, the SZPE leakage correction preserves the temperature dependence of the total rate coefficient and all individual product channels, while systematically increasing the rate coefficients at lower temperatures (by a factor of ~3 at 200 K) relative to the uncorrected values. This enhancement is consistent with the removal of a substantial fraction of non-reactive trajectories that violate the ZPE criterion, an effect particularly pronounced at low temperatures where interaction times are longer. The impact diminishes significantly at higher temperatures. In contrast, the RZPE correction yields a slight systematic decrease in both the total rate coefficient and the rates of all individual product channels across the entire temperature range relative to the NZPE values, as it selectively discards only reactive trajectories while retaining the full non-reactive ensemble. It is worth noting that the NZPE and RZPE rates are close to the theoretical results of Klippenstein *et al*.[11] at low temperatures (within a factor of 1.5) and agree well with the experimental value of Adamson *et al*.[14] at room temperature. By comparison with the absolute rates, the product branching ratios exhibit minimal sensitivity to the choice of ZPE treatment across all three schemes (Figure S3), indicating that relative product yields are robust with respect to ZPE handling. The strong agreement of our SZPE results with experimental data of Inomata and Washida[15] in the 200–500 K range further supports the physical reasonableness of this symmetric approach for the $NH_2$ + O system.

It is important to note that this study focuses exclusively on the ground doublet ( $^2A'$) potential energy surface. The reaction may, in principle, also proceed on low-lying quartet surfaces ( $^4A''$ and $^4A'$), which correlate to the same $\mathrm{NH_2(^2B_1) + O(^3P)}$ asymptote. Previous high-level theoretical studies have characterized these quartet surfaces. Yang *et al.*[19] reported a barrier of 6.7 kcal/mol for the direct abstraction reaction on the $^4A'$ surface at the G2 level, while Duan and Page[22] obtained barriers of 8.9 and 9.3 kcal/mol on the $^4A''$and $^4A'$ surfaces, respectively, at the MRCI/cc-pVTZ level. Based on variational transition state theory calculations including tunneling, Duan and Page[22] concluded that the abstraction reaction on the quartet surfaces begins to compete with the addition/isomerization/dissociation pathways only at temperatures above approximately 1000–1500 K. Consistent with these findings, the absence of these surfaces in our dynamics model may lead to a slight underestimation of the total rate coefficient at the highest temperatures considered here (e.g., $T$ > 1500 K), particularly for the NH + OH channel. Future work should aim to construct and include these quartet surfaces for a complete description of the high-temperature kinetics.

## IV. Conclusions

In this work, we have constructed a global full-dimensional potential energy surface for the ground $^2A'$ state of the $NH_2$ + O reaction based on approximately 62,000 high-level *ab initio* points computed at the MRCI-F12/cc-pVTZ-F12 level of theory. The PES, developed using the PIP-NN approach, demonstrates high fidelity in reproducing stationary points, reaction pathways, and reaction energetics across all accessible product channels. Extensive QCT calculations on this PES have yielded thermal rate coefficients and product branching ratios over a broad temperature range (200-2500 K) providing a first-principles kinetic dataset for this system under combustion-relevant conditions. Our results confirm the dominance of the HNO + H channel and the growing importance of the NH + OH channel as temperature rises, along with a non-negligible, temperature-insensitive contribution from the NO + $H_2$ channel. The theoretically validated rate predictions for the $NH_2$ + O reaction are expected to significantly improve the predictive accuracy of simulations concerning $NH_3$ ignition and oxidation processes. The kinetics data and mechanistic insights presented here are therefore critical for refining detailed kinetic models of ammonia combustion.

**Acknowledgements**

This work was supported by the National Natural Science Foundation of China (Grant No. 22503046 to J.Z. and Grant No. 12274229 to W.H.).

**Data availability statement**

The Supplemental Information contains complementary multireference diagnostics ($T_1$, $D_1$, and $D_2$) and optimized Cartesian coordinates for all identified stationary points (reactants, products, transition states, and intermediates) on the $NH_2O(^2A')$ PES. A detailed comparison of thermal rate coefficients and product branching ratios for the $NH_2$ + O reaction under three different zero-point energy treatments (SZPE, RZPE, and NZPE) is also included. Further supporting data are available from the corresponding author upon reasonable request.

## References

[1] A. Valera-Medina, H. Xiao, M. Owen-Jones, W. I. F. David, and P. J. Bowen, Ammonia for power, Prog. Energy Combust. Sci. **69**, 63–102 (2018).
[2] H. Kobayashi, A. Hayakawa, K. K. A. Somarathne, and E. C. Okafor, Science and technology of ammonia combustion, Proc. Combust. Inst. **37**, 109–133 (2019).
[3] O. Elishav, B. Mosevitzky Lis, E. M. Miller, D. J. Arent, A. Valera-Medina, A. Grinberg Dana, G. E. Shter, and G. S. Grader, Progress and Prospective of Nitrogen-Based Alternative Fuels, Chem. Rev. **120**, 5352–5436 (2020).
[4] A. Valera-Medina, F. Amer-Hatem, A. K. Azad, I. C. Dedoussi, M. de Joannon, R. X. Fernandes, P. Glarborg, H. Hashemi, X. He, S. Mashruk, J. McGowan, C. Mounaim-Rouselle, A. Ortiz-Prado, A. Ortiz-Valera, I. Rossetti, B. Shu, M. Yehia, H. Xiao, and M. Costa, Review on Ammonia as a Potential Fuel: From Synthesis to Economics, Energy Fuels **35**, 6964–7029 (2021).
[5] L. Kang, W. Pan, J. Zhang, W. Wang, and C. Tang, A review on ammonia blends combustion for industrial applications, Fuel **332**, 126150 (2023).
[6] H. Nakamura, S. Hasegawa, and T. Tezuka, Kinetic modeling of ammonia/air weak flames in a micro flow reactor with a controlled temperature profile, Combust. Flame **185**, 16–27 (2017).
[7] B. Mei, X. Zhang, S. Ma, M. Cui, H. Guo, Z. Cao, and Y. Li, Experimental and kinetic modeling investigation on the laminar flame propagation of ammonia under oxygen enrichment and elevated pressure conditions, Combust. Flame **210**, 236–246 (2019).
[8] X. Han, Z. Wang, Y. He, Y. Zhu, and K. Cen, Experimental and kinetic modeling study of laminar burning velocities of $NH_3$/syngas/air premixed flames, Combust. Flame **213**, 1–13 (2020).
[9] A. Stagni, C. Cavallotti, S. Arunthanayothin, Y. Song, O. Herbinet, F. Battin-Leclerc, and T. Faravelli, An experimental, theoretical and kinetic-modeling study of the gas-phase oxidation of ammonia, React. Chem. Eng. **5**, 696–711 (2020).
[10] X. Han, M. L. Lavadera, and A. A. Konnov, An experimental and kinetic modeling study on the laminar burning velocity of $NH_3$+$N_2O^+$ air flames, Combust. Flame **228**, 13–28 (2021).
[11] S. J. Klippenstein, C. R. Mulvihill, and P. Glarborg, Theoretical Kinetics Predictions for Reactions on the NH2O Potential Energy Surface, J. Phys. Chem. A **127**, 8650–8662 (2023).
[12] M. Gehring, K. Hoyermann, H. Schacke, and J. Wolfrum, Direct studies of some elementary steps for the formationand destruction of nitric oxide in the H-N-O system, Symp. (Int.) Combust. **14**, 99–105 (1973).
[13] P. Dransfeld, W. Hack, H. Kurzke, F. Temps, and H. G. Wagner, Direct studies of elementary reactions of $NH_2$-radicals in the gas phase, Chem. Phys. Lett. **20**, 655–663 (1985).
[14] J. D. Adamson, S. K. Farhat, C. L. Morter, G. P. Glass, R. F. Curl, and L. F. Phillips, The Reaction of $NH_2$ with O, J. Phys. Chem. **98**, 5665–5669 (1994).
[15] S. Inomata, and N. Washida, Rate Constants for the Reactions of $NH_2$ and HNO with Atomic Oxygen at Temperatures between 242 and 473 K, J. Phys. Chem. A **103**, 5023–5031 (1999).
[16] D. Patel-Misra, D. G. Sauder, and P. J. Dagdigian, Internal state distribution of OD produced from the $O(^3P)$+$ND_2$ reaction, J. Chem. Phys. **95**, 955–962 (1991).
[17] D. Patel-Misra, and P. J. Dagdigian, Dynamics of the $O(^3P)$+$NH_2$ reaction: the HNO + H product channel, Chem. Phys. Lett. **185**, 387–392 (1991).
[18] S. P. Walch, Theoretical characterization of the reaction $NH_2$+O→products, J. Chem. Phys. **99**, 3804–3808 (1993).
[19] D. L. Yang, M. L. Koszykowski, and J. L. Durant, Jr., The reaction of $NH_2(X\,^2B_1)$ with O $(X\,^3P)$: A theoretical study employing Gaussian 2 theory, J. Chem. Phys. **101**, 1361–1368 (1994).
[20] M. Wolf, D. L. Yang, and J. L. Durant, Kinetic studies of NHx radical reactions, J. Photochem. Photobiol. A Chem. **80**, 85–93 (1994).
[21] J. Bozzelli, and A. Dean, Energized complex quantum Rice-Ramsperger-Kassel analysis of reactions of $NH_2$, $O_2$, and O atoms, J. Phys. Chem. **93**, 1058–1065 (1989).
[22] X. Duan, and M. Page, *Ab initio* variational transition state theory calculations for the O+$NH_2$ hydrogen abstraction reaction on the $^4A'$ and $^4A''$ potential energy surfaces, J. Chem. Phys. **102**, 6121–6127 (1995).

[23] R. Sumathi, D. Sengupta, and M. T. Nguyen, Theoretical Study of the $H_2$ + NO and Related Reactions of [$H_2NO$] Isomers, J. Phys. Chem. A **102**, 3175–3183 (1998).
[24] M. R. Soto, and M. Page, *Ab initio* variational transition-state-theory reaction-rate calculations for the gas-phase reaction H+HNO→$H_2$+NO, J. Chem. Phys. **97**, 7287–7296 (1992).
[25] M. Page, and M. R. Soto, Radical addition to HNO. *Ab initio* reaction path and variational transition state theory calculations for H+HNO→$H_2NO$ and H+HNO→HNOH, J. Chem. Phys. **99**, 7709–7717 (1993).
[26] H. M. T. Nguyen, S. Zhang, J. Peeters, T. N. Truong, and M. T. Nguyen, Direct ab initio dynamics studies of the reactions of HNO with H and OH radicals, Chem. Phys. Lett. **388**, 94–99 (2004).
[27] E. Vessally, S. Ebrahimi, M. Goodarzi, and A. Seif, Insight into detailed mechanism of the atmospheric reaction of imidogen with hydroxyl: a computational study, Struct. Chem. **25**, 169–175 (2014).
[28] S. S. Asemani, and S. H. Mousavipour, Dynamics of imidogen reaction with hydroxyl radical: a theoretical approach, J. Iran. Chem. Soc. **17**, 1987–2000 (2020).
[29] R. Sharafdini, and S. Ramazani, Dynamic and Kinetic Parameters and Energy Exchanges of Particles in Reaction of NH + OH and Deuterated Analogues on an Interpolated Potential Energy Surface, ChemistrySelect **5**, 3518–3528 (2020).
[30] Y. Kurosaki, and T. Takayanagi, *Ab initio* molecular orbital study of potential energy surface for the $H_2NO(^2B_1)$→$NO(^2\Pi)$+$H_2$ reaction, J. Mol. Struct.: THEOCHEM **507**, 119–126 (2000).
[31] Z. Homayoon, and J. M. Bowman, A Global Potential Energy Surface Describing the $N(^2D)$ + $H_2O$ Reaction and a Quasiclassical Trajectory Study of the Reaction to NH + OH, J. Phys. Chem. A **118**, 545–553 (2014).
[32] M. Isegawa, F. Liu, S. Maeda, and K. Morokuma, Complete active space second order perturbation theory (CASPT2) study of $N(^2D)$ + $H_2O$ reaction paths on $D_1$ and $D_0$ potential energy surfaces: Direct and roaming pathways, J. Chem. Phys. **141**(2014).
[33] P. J. Knowles, and H.-J. Werner, An efficient method for the evaluation of coupling coefficients in configuration interaction calculations, Chem. Phys. Lett. **145**, 514–522 (1988).
[34] T. Shiozaki, G. Knizia, and H.-J. Werner, Explicitly correlated multireference configuration interaction: MRCI-F12, J. Chem. Phys. **134**(2011).
[35] J. Li, B. Jiang, and H. Guo, Permutation invariant polynomial neural network approach to fitting potential energy surfaces. II. Four-atom systems, J. Chem. Phys. **139**, 204103 (2013).
[36] B. Jiang, and H. Guo, Permutation invariant polynomial neural network approach to fitting potential energy surfaces, J. Chem. Phys. **139**, 054112 (2013).
[37] O. Tishchenko, J. Zheng, and D. G. Truhlar, Multireference Model Chemistries for Thermochemical Kinetics, J. Chem. Theory Comput. **4**, 1208–1219 (2008).
[38] M. P. Deskevich, D. J. Nesbitt, and H.-J. Werner, Dynamically weighted multiconfiguration self-consistent field: Multistate calculations for F+$H_2O$→HF+OH reaction paths, J. Chem. Phys. **120**, 7281–7289 (2004).
[39] R. Dawes, A. W. Jasper, C. Tao, C. Richmond, C. Mukarakate, S. H. Kable, and S. A. Reid, Theoretical and Experimental Spectroscopy of the $S_2$ State of CHF and CDF: Dynamically Weighted Multireference Configuration Interaction Calculations for High-Lying Electronic States, J. Phys. Chem. Lett. **1**, 641–646 (2010).
[40] S. R. Langhoff, and E. R. Davidson, Configuration interaction calculations on the nitrogen molecule, Int. J. Quantum Chem. **8**, 61–72 (1974).
[41] K. A. Peterson, T. B. Adler, and H.-J. Werner, Systematically convergent basis sets for explicitly correlated wavefunctions: The atoms H, He, B–Ne, and Al–Ar, J. Chem. Phys. **128**, 084102 (2008).
[42] G. Knizia, T. B. Adler, and H.-J. Werner, Simplified CCSD(T)-F12 methods: theory and benchmarks, J. Chem. Phys. **130**, 221106 (2009).
[43] T. B. Adler, G. Knizia, and H.-J. Werner, A simple and efficient CCSD(T)-F12 approximation, J. Chem. Phys. **127**, 221106 (2007).
[44] H.-J. Werner, P. J. Knowles, G. Knizia, F. R. Manby, and M. Schütz, Molpro: a general-purpose quantum chemistry program package, WIREs Comput. Mol. Sci. **2**, 242–253 (2012).

[45] B. Jiang, J. Li, and H. Guo, Potential energy surfaces from high fidelity fitting of ab initio points: the permutation invariant polynomial - neural network approach, Int. Rev. Phys. Chem. **35**, 479–506 (2016).
[46] B. J. Braams, and J. M. Bowman, Permutationally invariant potential energy surfaces in high dimensionality, Int. Rev. Phys. Chem. **28**, 577–606 (2009).
[47] M. T. Hagan, and M. B. Menhaj, Training feedforward networks with the Marquardt algorithm, IEEE Trans. Neural Netw. **5**, 989–993 (1994).
[48] L. M. Raff, R. Komanduri, M. Hagan, and S. T. S. Bukkapatnam, *Neural Networks in Chemical Reaction Dynamics* (Oxford University Press, 2012),
[49] W. L. Hase, R. J. Duchovic, X. Hu, A. Komornicki, K. F. Lim, D.-h. Lu, G. H. Peslherbe, K. N. Swamy, S. Linde, and A. Varandas, A general chemical dynamics computer program, Quantum Chem. Program Exch. Bull **16**, 671 (1996).
[50] Y. Guo, D. L. Thompson, and T. D. Sewell, Analysis of the zero-point energy problem in classical trajectory simulations, J. Chem. Phys. **104**, 576–582 (1996).
[51] D. G. Truhlar, Multiple potential energy surfaces for reactions of species in degenerate electronic states, J. Chem. Phys. **56**, 3189–3190 (1972).
[52] K. P. Huber, and G. Herzberg, *Molecular spectra and molecular structure: IV. Constants of diatomic molecules* (Springer Science & Business Media, 2013),
[53] S. J. Klippenstein, L. B. Harding, B. Ruscic, R. Sivaramakrishnan, N. K. Srinivasan, M. C. Su, and J. V. Michael, Thermal decomposition of $NH_2OH$ and subsequent reactions: *Ab initio* transition state theory and reflected shock tube experiments, J. Phys. Chem. A **113**, 10241–10259 (2009).
[54] J. A. Miller, M. D. Smooke, R. M. Green, and R. J. Kee, Kinetic Modeling of the Oxidation of Ammonia in Flames, Combust. Sci. Technol. **34**, 149–176 (1983).
[55] A. M. Dean, and J. W. Bozzelli, in *Gas-phase combustion chemistry* (Springer, 2000), pp. 125–341.

TABLE I. M diagnostics of all stationary points (see structures in Figure 3) on the computed $NH_2O(^2A')$ PES.

| Species | M diagnostic |
|---|---|
| $NH_2(^2B_1) + O(^3P)$ | 0.679 |
| $HNO(^1A'') + H$ | 0.617 |
| $NH(^3\Sigma^-) + OH(^2\Pi)$ | 0.385 |
| $NO(^2\Pi) + H_2(^1\Sigma_g^+)$ | 0.527 |
| $HON(^3A'') + H$ | 0.326 |
| $H_2NO$ | 0.390 |
| *trans*-HNOH | 0.254 |
| *cis*-HNOH | 0.247 |
| TS1 | 0.437 |
| TS2 | 0.415 |
| TS3 | 0.419 |
| TS4 | 0.578 |
| TS5 | 0.589 |

Table II. Geometrical parameters (distances in Å and angles in °) of optimized asymptotes and stationary points (see structures in Figure 3) optimized at the MRCI-F12/cc-pVTZ-F12 level of theory for the $NH_2$ + O($^3$P) reaction.

| Species | $R_{\mathrm{NH}}$ | $R_{\mathrm{NH'}}$ | $R_{\mathrm{OH}}$ | $R_{\mathrm{NO}}$ | $R_{\mathrm{HH}}$ | $\theta_{\mathrm{HNH}}$ | $\theta_{\mathrm{HNO}}$ | $\theta_{\mathrm{H'NO}}$ | $\theta_{\mathrm{NOH}}$ | $\varphi_{\mathrm{HNOH}}$ |
|---|---|---|---|---|---|---|---|---|---|---|
| $NH_2$ | 1.024 | 1.024 | / | / | / | 102.88 | / | / | / | / |
| HNO | 1.041 | / | / | 1.211 | / | / | 108.18 | / | / | / |
| HON | / | / | 0.968 | 1.328 | / | / | / | / | 106.89 | / |
| NH | 1.036 | / | / | / | / | / | / | / | / | / |
| OH | / | / | 0.969 | / | / | / | / | / | / | / |
| NO | / | / | / | 1.152 | / | / | / | / | / | / |
| $H_2$ | / | / | / | / | 0.743 | / | / | / | / | / |
| $H_2NO$ | 1.003 | 1.012 | / | 1.274 | / | 119.23 | 118.51 | 118.75 | / | 158.72 |
| *trans*-HNOH | 1.016 | / | 0.962 | 1.371 | / | / | 100.81 | / | 103.42 | 180.00 |
| *cis*-HNOH | 1.018 | / | 0.964 | 1.370 | / | / | 105.68 | / | 109.56 | 0.00 |
| TS1 | 1.019 | 1.191 | / | 1.413 | / | 128.25 | 106.29 | 54.41 | / | 125.74 |
| TS2 | 1.036 | / | 0.978 | 1.410 | / | / | 102.80 | / | 105.40 | 85.40 |
| TS3 | 1.281 | 1.281 | / | 1.236 | / | 45.37 | 108.50 | 108.50 | / | 47.99 |
| TS4 | 1.034 | / | 1.746 | 1.236 | / | 121.35 | 105.41 | 44.40 | / | 119.32 |
| TS5 | 1.061 | 2.250 | / | 1.206 | 2.554 | 94.03 | 109.74 | 135.39 | / | 118.03 |

TABLE III. Calculated energies (kcal/mol, relative to the $NH_2$ + O($^3$P) asymptote) and harmonic vibrational frequencies (cm$^{-1}$) of all stationary points (see structures in Figure 3) on the $NH_2O(^2A')$ PES. Values in parentheses include zero-point energy corrections.

| Species | Methods | *E* (kcal/mol) | Harmonic frequencies (cm$^{-1}$) | | | | | |
|---|---|---|---|---|---|---|---|---|
| | | | 1 | 2 | 3 | 4 | 5 | 6 |
| $NH_2$ + O | PES[a] | 0 | 1542.55 | 3374.15 | 3456.16 | / | / | / |
| | MRCI[b] | 0 | 1553.26 | 3380.22 | 3474.87 | / | / | / |
| | UCCSD(T)[c] | 0 | 1545.13 | 3361.50 | 3455.10 | / | / | / |
| | Expt.[d] | 0 | 1497.33 | 3219.47 | 3325.00 | / | / | / |
| HNO + H | PES[a] | -23.13(-26.09) | 1572.04 | 1752.26 | 2952.34 | / | / | / |
| | MRCI[b] | -22.98(-26.06) | 1558.48 | 1615.60 | 3127.47 | / | / | / |
| | UCCSD(T)[c] | -22.49(-23.48) | 1547.39 | 1605.42 | 2956.62 | / | / | / |
| | Expt.[d] | (-26.24 ± 0.05) | 1500.82 | 1565.33 | 2684.00 | / | / | / |
| NH + OH | PES[a] | -7.40(-9.36) | 3282.22 | 3768.17 | / | / | / | / |
| | MRCI[b] | -7.28(-9.26) | 3284.16 | 3752.66 | / | / | / | / |
| | UCCSD(T)[c] | -7.61(-9.18) | 3265.72 | 3719.69 | / | / | / | / |
| | Expt.[d] | (-9.50 ± 0.07) | 3282.23 | 3737.74 | / | / | / | / |
| NO + $H_2$ | PES[a] | -79.64(-82.63) | 1902.25 | 4353.29 | / | / | / | / |
| | MRCI[b] | -79.25(-82.49) | 1911.21 | 4382.90 | / | / | / | / |
| | UCCSD(T)[c] | -78.84(-79.58) | 2116.07 | 4396.76 | / | / | / | / |
| | Expt.[d] | (-82.49 ± 0.04) | 1904.20 | 4401.21 | / | / | / | / |
| HON + H | PES[a] | 3.44(0.16) | 1097.17 | 1215.88 | 3722.90 | / | / | / |
| | MRCI[b] | 3.35(0.07) | 1132.42 | 1253.02 | 3726.23 | / | / | / |
| | UCCSD(T)[c] | 3.12(1.33) | 1285.44 | 1473.64 | 3314.05 | / | / | / |
| | Expt.[d] | (-0.43 ± 0.21) | / | / | / | / | / | / |
| $H_2NO$ | PES[a] | -90.28(-85.58) | 145.39 | 1275.93 | 1410.45 | 1748.99 | 3498.46 | 3647.73 |
| | MRCI[b] | -89.43(-84.42) | 386.64 | 1297.23 | 1372.84 | 1681.63 | 3485.97 | 3688.66 |
| *trans*-HNOH | PES[a] | -83.74(-78.62) | 715.17 | 1104.54 | 1245.97 | 1540.74 | 3388.29 | 3833.69 |
| | MRCI[b] | -83.32(-78.08) | 750.96 | 1099.30 | 1289.44 | 1600.40 | 3522.74 | 3808.60 |
| *cis*-HNOH | PES[a] | -77.78(-73.40) | 527.92 | 1095.81 | 1329.92 | 1479.07 | 3353.69 | 3765.78 |
| | MRCI[b] | -77.66(-72.81) | 577.73 | 1087.57 | 1355.69 | 1537.65 | 3472.28 | 3767.58 |
| TS1 | PES[a] | -36.99(-34.09) | 2044.23i | 770.01 | 1003.49 | 1378.05 | 2628.98 | 3436.14 |
| | MRCI[b] | -36.75(-35.54) | 2146.75i | 804.60 | 1023.85 | 1372.77 | 2592.09 | 3463.61 |
| TS2 | PES[a] | -69.95(-66.68) | 951.68i | 985.55 | 1228.99 | 1450.13 | 3310.88 | 3800.00 |
| | MRCI[b] | -68.61(-65.48) | 863.83i | 971.16 | 1272.98 | 1479.33 | 3271.93 | 3597.58 |
| TS3 | PES[a] | -23.55(-25.72) | 1526.02i | 540.89 | 1131.90 | 1216.22 | 1473.63 | 2490.76 |
| | MRCI[b] | -22.74(-25.00) | 1784.82i | 834.33 | 1136.29 | 1144.81 | 1428.79 | 2278.38 |
| TS4 | PES[a] | -16.44(-18.77) | 1230.90i | 288.02 | 465.29 | 1420.28 | 1552.34 | 3021.24 |
| | MRCI[b] | -16.37(-18.28) | 988.74i | 343.07 | 496.67 | 1412.76 | 1589.68 | 3229.13 |
| TS5 | PES[a] | -22.61(-24.95) | 461.46i | 245.12 | 465.66 | 1443.88 | 1568.10 | 2944.90 |
| | MRCI[b] | -22.15(-24.54) | 333.65i | 228.63 | 431.33 | 1578.88 | 1600.79 | 2898.88 |

[a] Fitted PES, this work; [b] MRCI-F12/cc-pVTZ-F12, this work; [c] UCCSD(T)-F12a/cc-pVTZ-F12; [d] Experimental values.[52] Experimental anharmonic fundamentals from Ref.[52] are listed for qualitative comparison to validate the PES topology.

Table IV. Rate coefficients ($cm^3 \cdot molecule^{-1} \cdot s^{-1}$) for the $NH_2$ + $O(^3P)$ reaction at different temperatures from QCT calculations. $k$ is the total rate coefficient. $k_{HNO+H}$, $k_{NH+OH}$, $k_{NO+H_2}$, and $k_{HON+H}$ are channel-specific rates.

| $T$ / K | $k$ | $k_{HNO+H}$ | $k_{NH+OH}$ | $k_{NO+H_2}$ | $k_{HON+H}$ |
|---|---|---|---|---|---|
| 200 | $2.02\times10^{-10}$ | $1.42\times10^{-10}$ | $3.03\times10^{-11}$ | $2.90\times10^{-11}$ | $1.19\times10^{-12}$ |
| 298 | $1.63\times10^{-10}$ | $1.13\times10^{-10}$ | $2.58\times10^{-11}$ | $2.29\times10^{-11}$ | $9.84\times10^{-13}$ |
| 500 | $1.04\times10^{-10}$ | $7.11\times10^{-11}$ | $1.84\times10^{-11}$ | $1.40\times10^{-11}$ | $7.32\times10^{-13}$ |
| 700 | $7.58\times10^{-11}$ | $5.02\times10^{-11}$ | $1.51\times10^{-11}$ | $9.77\times10^{-12}$ | $6.61\times10^{-13}$ |
| 1000 | $5.50\times10^{-11}$ | $3.46\times10^{-11}$ | $1.28\times10^{-11}$ | $6.87\times10^{-12}$ | $6.92\times10^{-13}$ |
| 1500 | $4.10\times10^{-11}$ | $2.36\times10^{-11}$ | $1.20\times10^{-11}$ | $4.75\times10^{-12}$ | $7.19\times10^{-13}$ |
| 2000 | $3.52\times10^{-11}$ | $1.83\times10^{-11}$ | $1.21\times10^{-11}$ | $3.90\times10^{-12}$ | $8.68\times10^{-13}$ |
| 2500 | $3.31\times10^{-11}$ | $1.57\times10^{-11}$ | $1.29\times10^{-11}$ | $3.54\times10^{-12}$ | $9.38\times10^{-13}$ |

Table V. Temperature-dependent product branching ratios (in percent) for the $NH_2$ + $O(^3P)$ reaction. $f_{HNO+H}$, $f_{NH+OH}$, $f_{NO+H_2}$, and $f_{HON+H}$ denote the fractional yields of the respective channels.

| $T$ / K | $f_{HNO+H}$ | $f_{NH+OH}$ | $f_{NO+H_2}$ | $f_{HON+H}$ |
|---|---|---|---|---|
| 200 | 70.30 | 15.00 | 14.36 | 0.59 |
| 298 | 69.33 | 15.83 | 14.05 | 0.60 |
| 500 | 68.37 | 17.69 | 13.46 | 0.70 |
| 700 | 66.23 | 19.92 | 12.89 | 0.87 |
| 1000 | 62.91 | 23.27 | 12.49 | 1.26 |
| 1500 | 57.56 | 29.27 | 11.59 | 1.75 |
| 2000 | 51.99 | 34.38 | 11.08 | 2.47 |
| 2500 | 47.43 | 38.97 | 10.69 | 2.83 |

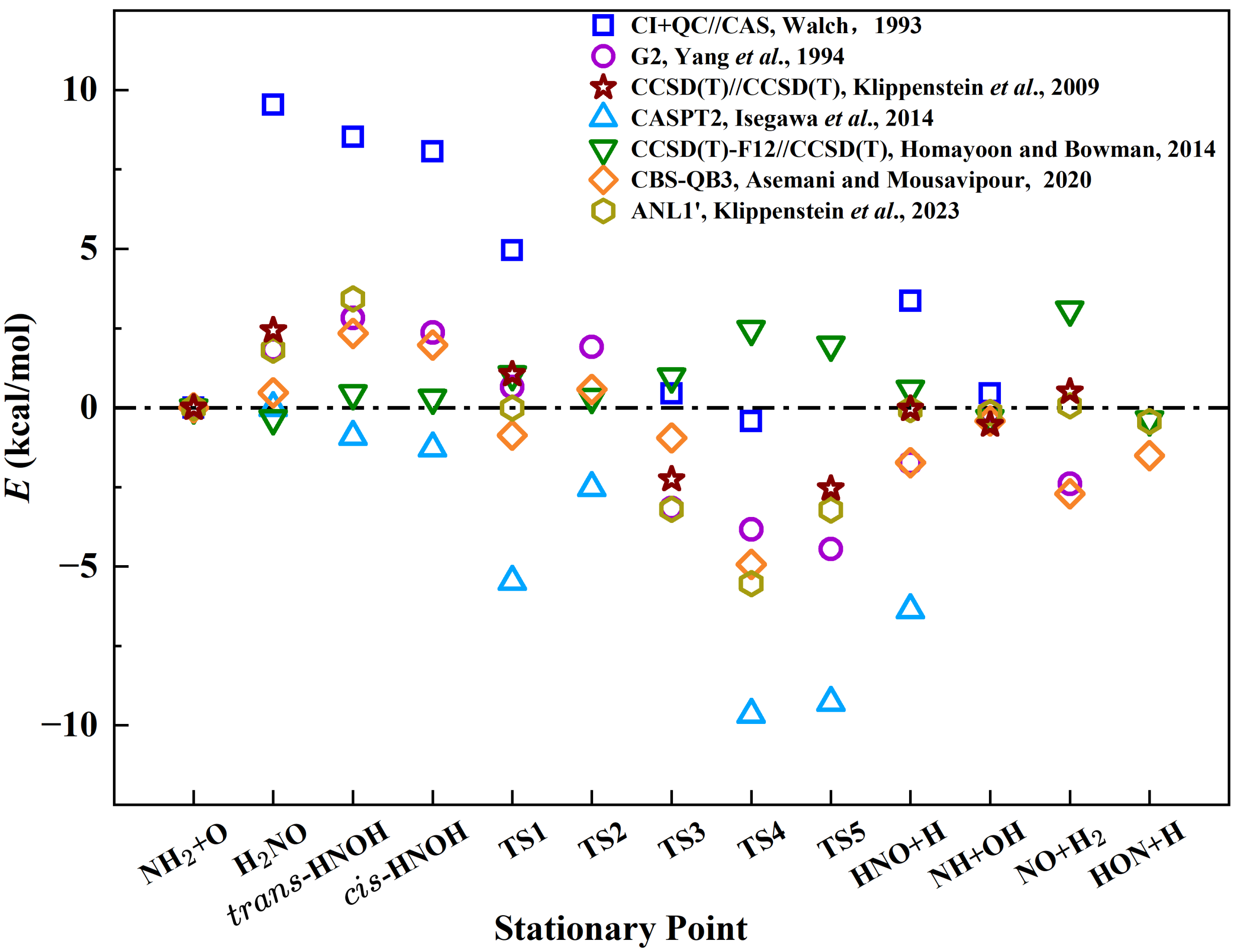

Fig. 1. Deviations of stationary point energies from the literature[11, 18, 19, 28, 31, 32, 53] benchmarked against the present MRCI-F12/cc-pVTZ-F12 calculations.

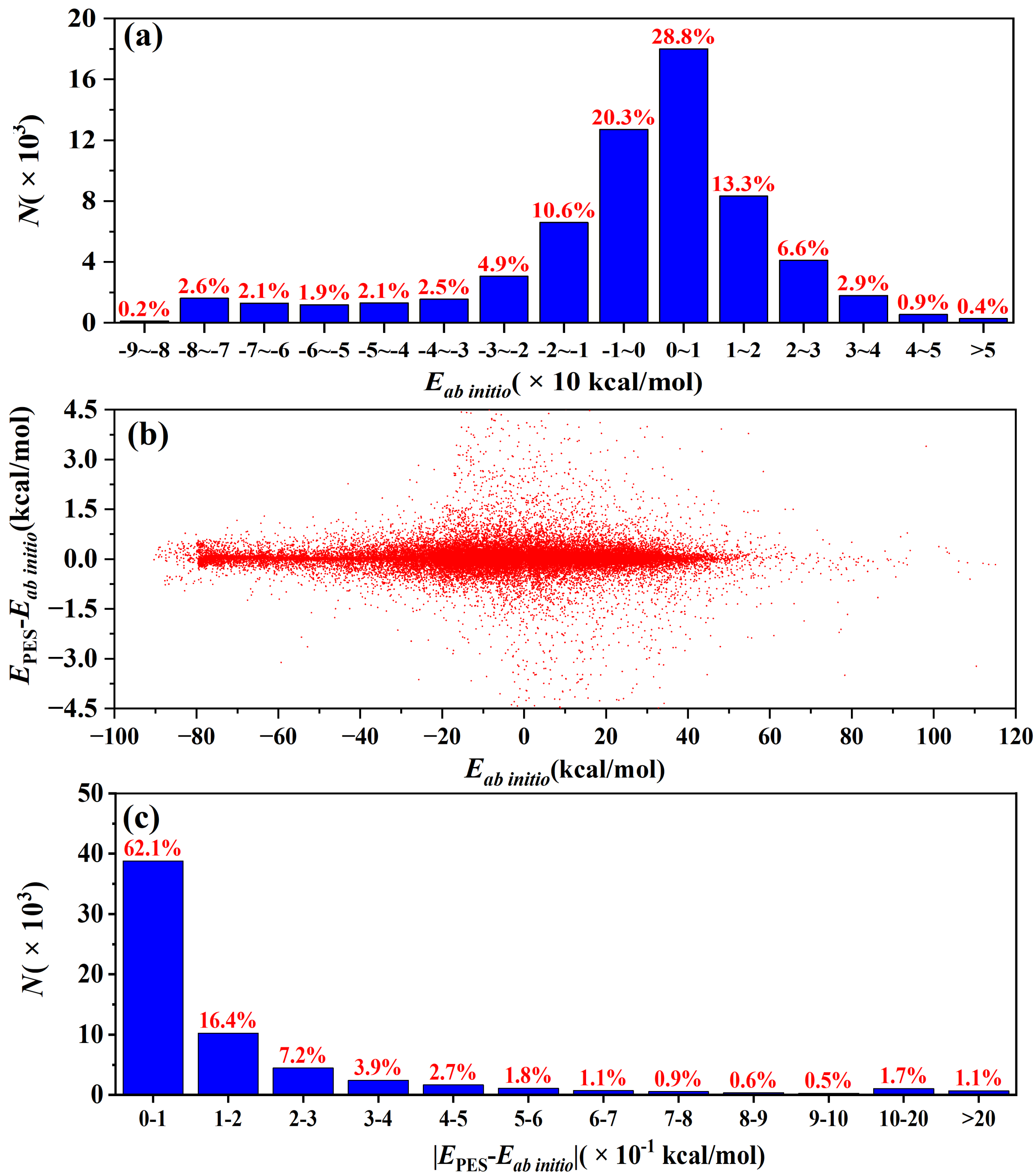

Fig. 2. Dataset and fitting accuracy of the $NH_2$ + O($^3$P) PIP-NN PES. (a) Energy distribution of the MRCI points used for fitting. (b) Fitting error ($E_{PES}$ -$E_{ab\ initio}$) versus MRCI reference energy ($E_{ab\ initio}$). (c) Histogram of the fitting errors. All energies are given relative to the $NH_2$ + O($^3$P) asymptote.

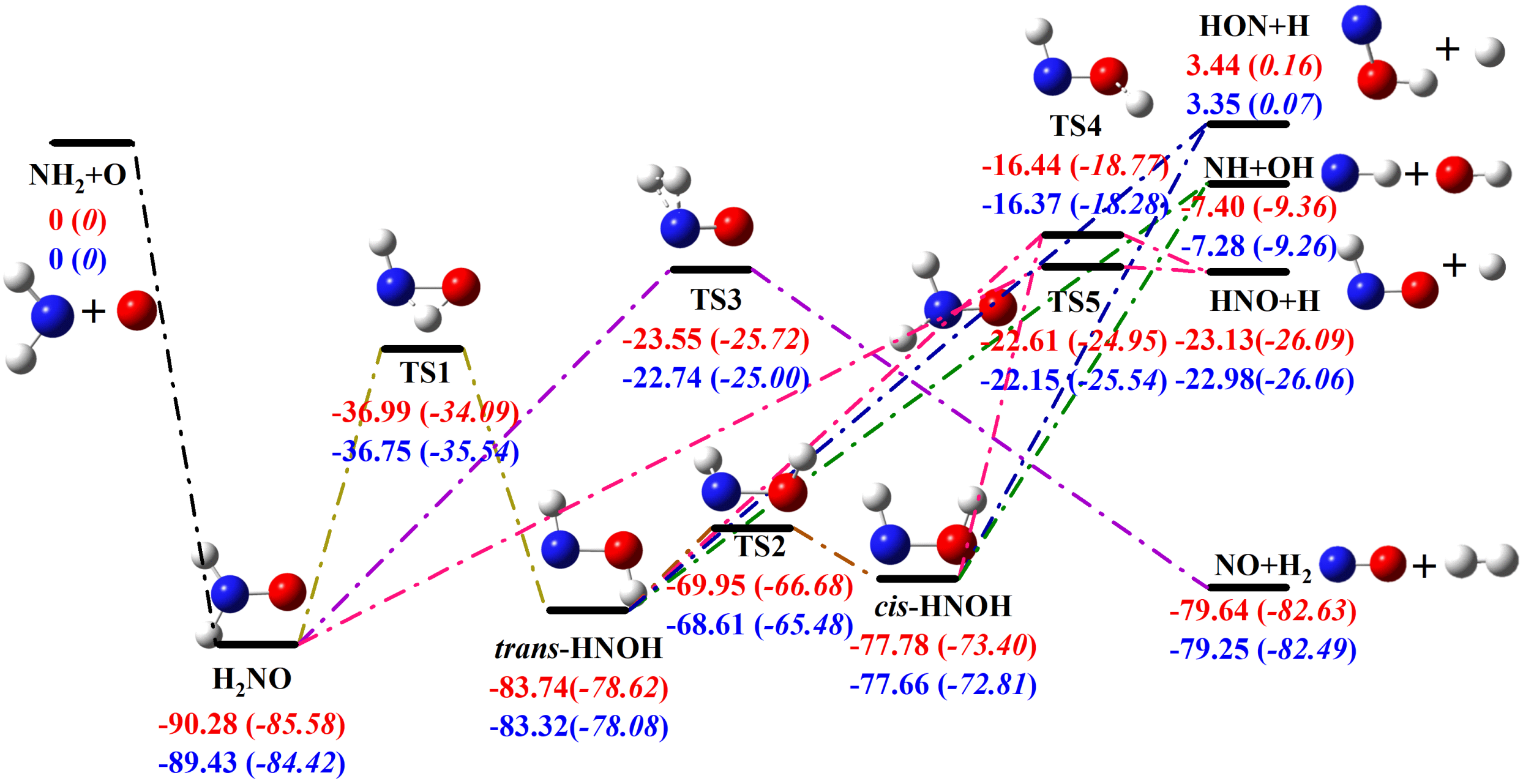

Fig. 3. Schematic energy diagram for the $NH_2$ + O($^3$P) reaction on the ground-state PES. Stationary-point energies (kcal/mol, relative to reactants) are shown with (italic, parentheses) and without (upright) zero-point correction, from the present PIP-NN PES (red) and direct MRCI calculations (blue).

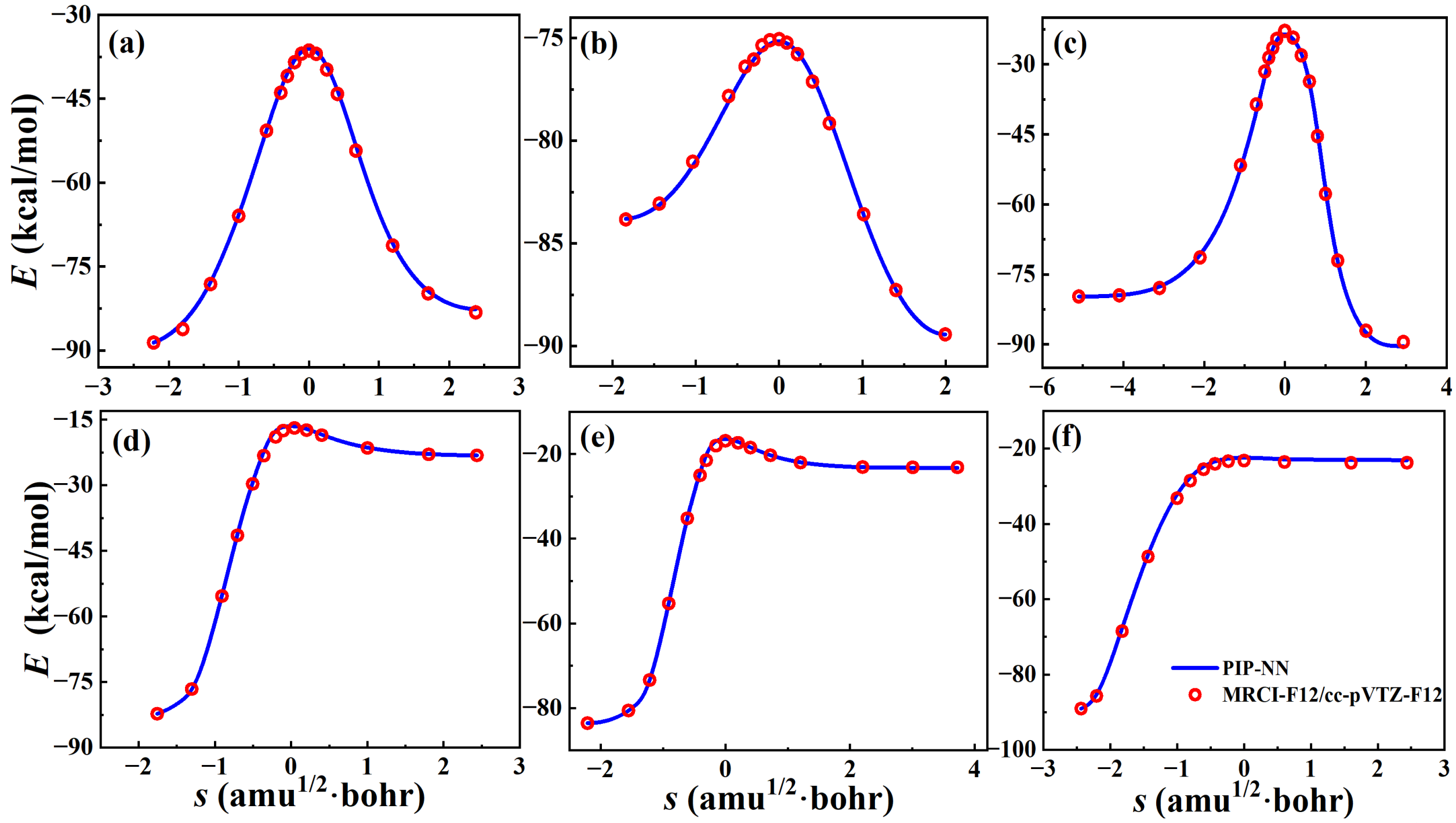

Fig. 4. Minimum energy paths along reaction coordinate for key processes in the $NH_2$ + O($^3$P) reaction. (a) $H_2NO$ → TS1 → *trans*-HNOH, (b) *trans*-HNOH → TS2 → *cis*-HNOH, (c) $H_2NO$→TS3→NO+$H_2$, (d) *trans*-HNOH→TS4→HNO+H, (e) *cis*-HNOH→TS4→HNO+H, and (f) $H_2NO$→TS5→HNO+H. Solid lines represent MEPs traced on the fitted PES, while symbols denote the benchmark MRCI energy points.

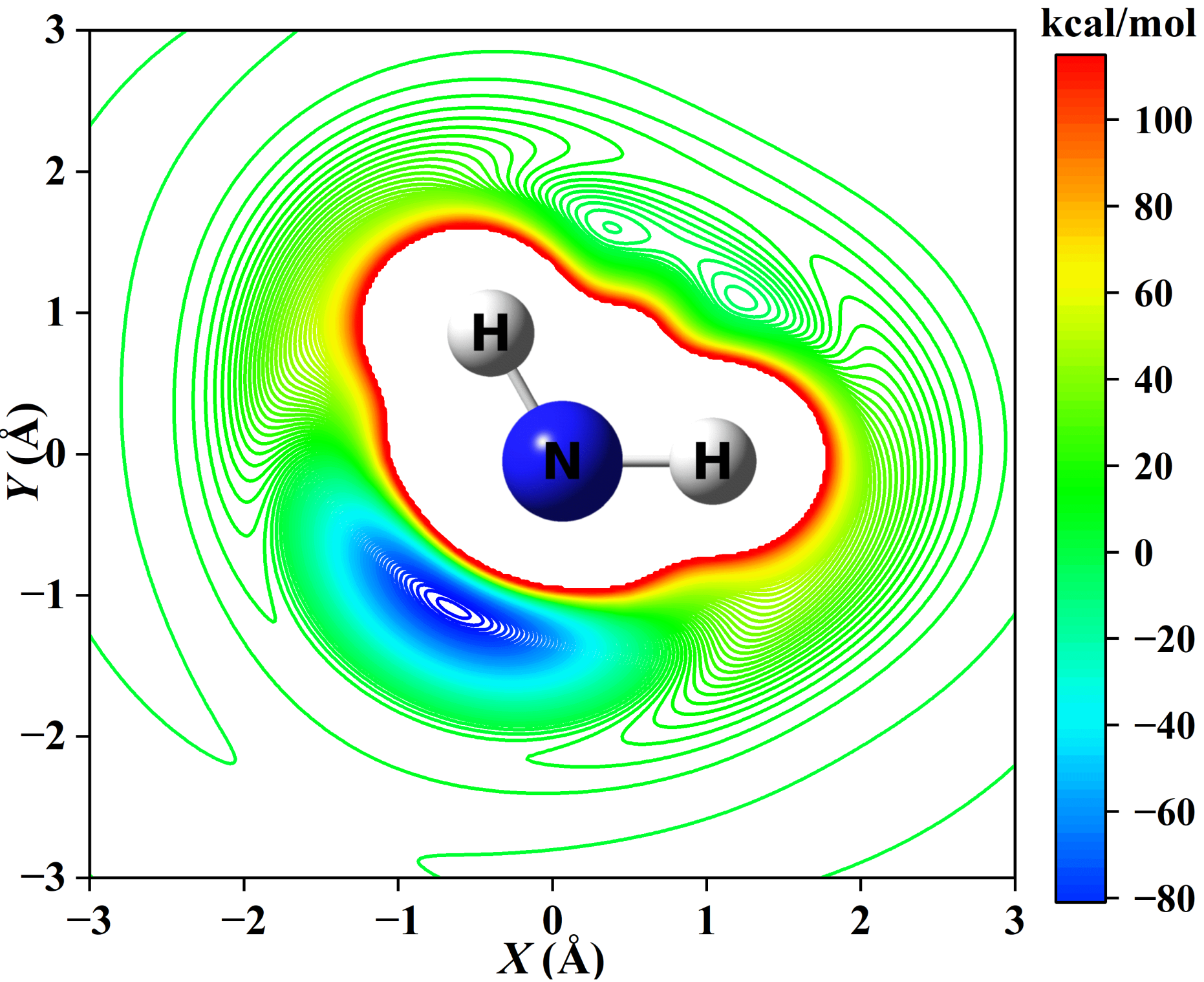

Fig. 5. Two-dimensional potential energy contour plot of the $NH_2O$ system. The $NH_2$ fragment is held fixed at its equilibrium geometry while the O atom is scanned across its plane. All energies (in kcal/mol) are relative to the $NH_2$ + O asymptote, with a contour interval of 2 kcal/mol.

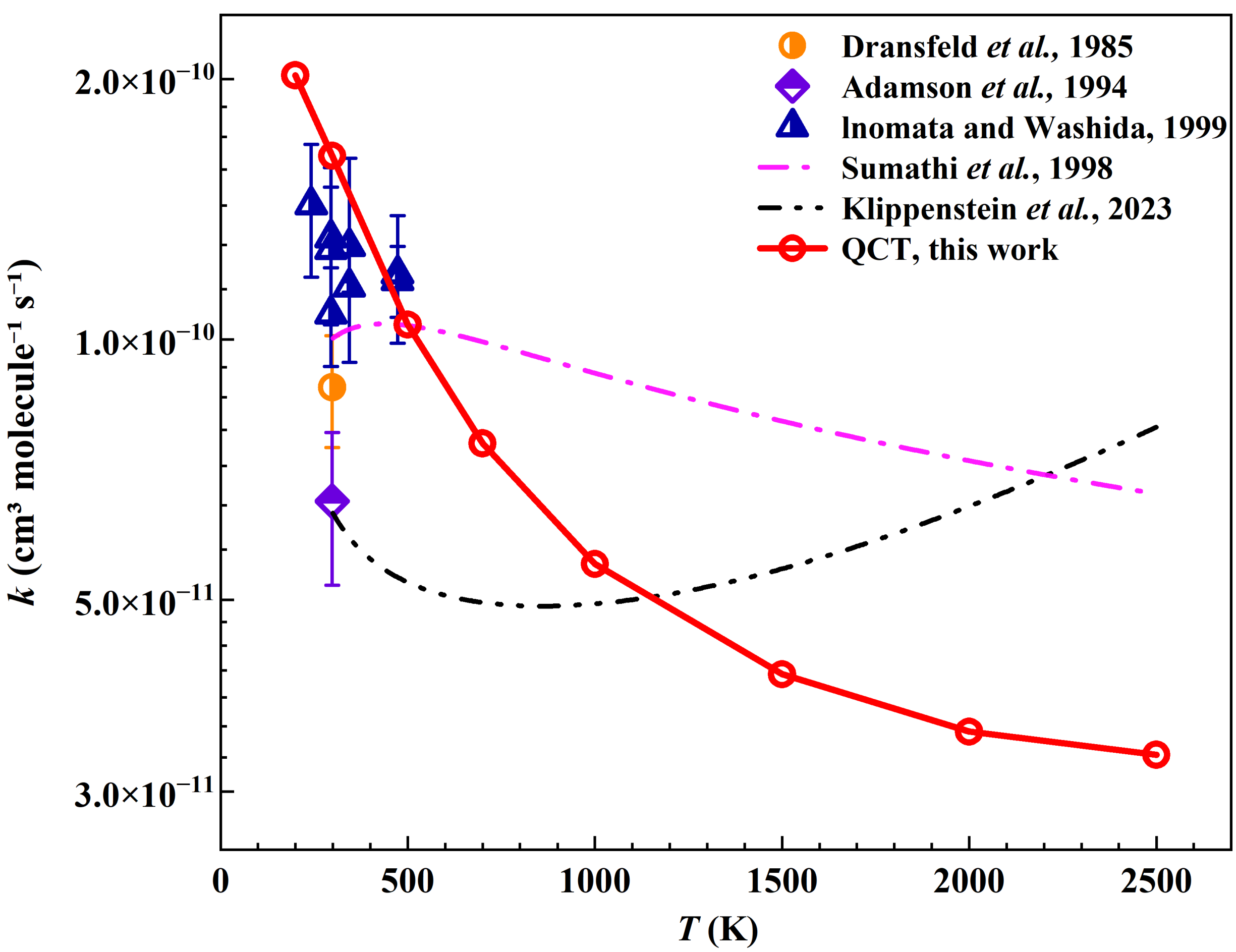

Fig. 6. Comparison of the total thermal rate coefficients for the $NH_2$ + O($^3$P) reaction as a function of temperature between the present QCT results and experimental[13-15] and theoretical[11, 23] data from the literature.

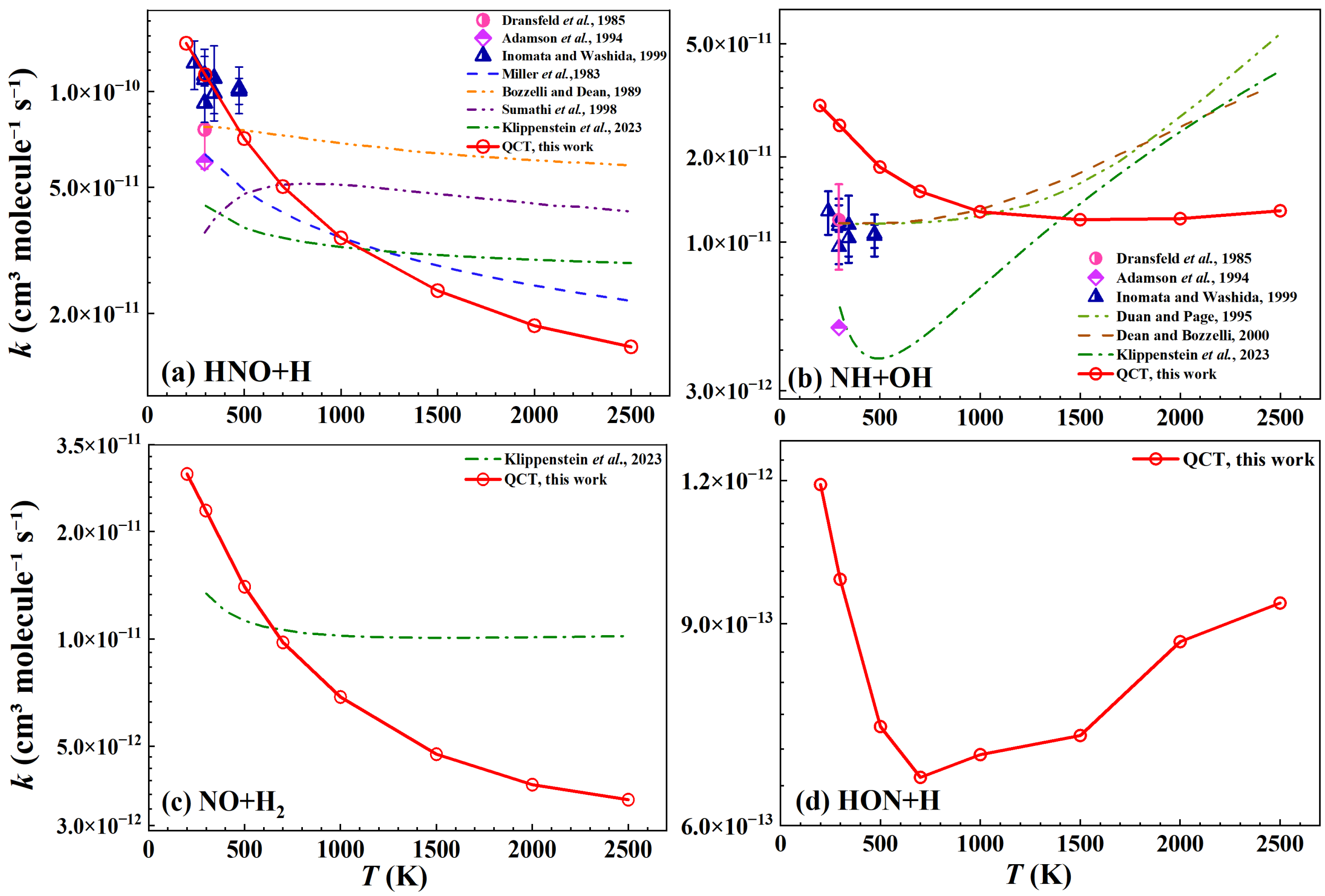

Fig. 7. Comparison of channel-specific thermal rate coefficients for the $NH_2$ + O($^3$P) reaction as a function of temperature. Panels (a)-(d) correspond to the HNO + H, NH + OH, NO + $H_2$, and HON + H channels, respectively, where the present QCT results are compared with experimental[13-15] and theoretical[11, 21-23, 54, 55] data from the literature.

Supplemental Information

# Clarifying $NH_2$ + O($^3$P) Reaction Dynamics: A Full-Dimensional MRCI, Machine-Learned PES Unravels High-Temperature Kinetics

Ying Xing, Weijie Hua, and Junxiang Zuo*

MIIT Key Laboratory of Semiconductor Microstructure and Quantum Sensing, Department of Applied Physics, School of Physics, Nanjing University of Science and Technology, Nanjing 210094, China

This PDF contains

*Corresponding author: jxzuo@njust.edu.cn.

## S1. Supplementary Tables and Figures

Table S1. Complementary multireference diagnostics ($T_1$, $D_1$, and $D_2$) for all stationary points on the $NH_2O(^2A')$ PES. Diagnostics were computed at the UCCSD(T)-F12a/cc-pVTZ-F12 level of theory, employing geometries optimized at the MRCI-F12/cc-pVTZ-F12 level.

| Species | $T_1$ diagnostic | $D_1$ diagnostic | $D_2$ diagnostic |
|---|---|---|---|
| $NH_2(^2B_1) + O(^3P)$ | 0.007 | 0.012 | 0.131 |
| $HNO(^1A'') + H$ | 0.015 | 0.043 | 0.203 |
| $NH(^3\Sigma^-) + OH(^2\Pi)$ | 0.008 | 0.014 | 0.153 |
| $NO(^2\Pi) + H_2(^1\Sigma_g^+)$ | 0.021 | 0.042 | 0.175 |
| $HON(^3A'') + H$ | 0.015 | 0.036 | 0.149 |
| $H_2NO$ | 0.019 | 0.058 | 0.161 |
| *trans*-HNOH | 0.017 | 0.052 | 0.148 |
| *cis*-HNOH | 0.017 | 0.053 | 0.147 |
| TS1 | 0.029 | 0.063 | 0.174 |
| TS2 | 0.014 | 0.037 | 0.156 |
| TS3 | 0.026 | 0.061 | 0.176 |
| TS4 | 0.058 | 0.146 | 0.222 |
| TS5 | 0.059 | 0.192 | 0.216 |

Table S2. Thermal rate coefficients (cm$^3$·molecule$^{-1}$·s$^{-1}$) for the $NH_2$ + O reaction from QCT simulations employing the RZPE correction scheme (QCT + RZPE).

| $T$ / K | $k^{\mathrm{RZPE}}$ | $k^{\mathrm{RZPE}}_{\mathrm{HNO+H}}$ | $k^{\mathrm{RZPE}}_{\mathrm{NH+OH}}$ | $k^{\mathrm{RZPE}}_{\mathrm{NO+H_2}}$ | $k^{\mathrm{RZPE}}_{\mathrm{HON+H}}$ |
|---|---|---|---|---|---|
| 200 | $5.46 \times 10^{-11}$ | $3.82 \times 10^{-11}$ | $8.19 \times 10^{-12}$ | $7.83 \times 10^{-12}$ | $3.21 \times 10^{-13}$ |
| 298 | $4.70 \times 10^{-11}$ | $3.26 \times 10^{-11}$ | $7.44 \times 10^{-12}$ | $6.61 \times 10^{-12}$ | $2.84 \times 10^{-13}$ |
| 500 | $3.93 \times 10^{-11}$ | $2.68 \times 10^{-11}$ | $6.94 \times 10^{-12}$ | $5.26 \times 10^{-12}$ | $2.76 \times 10^{-13}$ |
| 700 | $3.57 \times 10^{-11}$ | $2.37 \times 10^{-11}$ | $7.09 \times 10^{-12}$ | $4.60 \times 10^{-12}$ | $3.11 \times 10^{-13}$ |
| 1000 | $3.26 \times 10^{-11}$ | $2.05 \times 10^{-11}$ | $7.61 \times 10^{-11}$ | $4.07 \times 10^{-12}$ | $4.10 \times 10^{-13}$ |
| 1500 | $3.04 \times 10^{-11}$ | $1.74 \times 10^{-11}$ | $8.89 \times 10^{-12}$ | $3.51 \times 10^{-12}$ | $5.33 \times 10^{-13}$ |
| 2000 | $2.94 \times 10^{-11}$ | $1.53 \times 10^{-11}$ | $1.01 \times 10^{-11}$ | $3.26 \times 10^{-12}$ | $7.24 \times 10^{-13}$ |
| 2500 | $2.95 \times 10^{-11}$ | $1.40 \times 10^{-11}$ | $1.15 \times 10^{-11}$ | $3.16 \times 10^{-12}$ | $8.37 \times 10^{-13}$ |

Table S3. Thermal rate coefficients ($cm^3 \cdot molecule^{-1} \cdot s^{-1}$) for the $NH_2$ + O reaction from QCT simulations performed without enforcing ZPE conservation (QCT + NZPE).

| $T$ / K | $k^{\mathrm{NZPE}}$ | $k^{\mathrm{NZPE}}_{\mathrm{HNO+H}}$ | $k^{\mathrm{NZPE}}_{\mathrm{NH+OH}}$ | $k^{\mathrm{NZPE}}_{\mathrm{NO+H_2}}$ | $k^{\mathrm{NZPE}}_{\mathrm{HON+H}}$ |
|---|---|---|---|---|---|
| 200 | $6.06 \times 10^{-11}$ | $3.99 \times 10^{-11}$ | $1.18 \times 10^{-11}$ | $8.33 \times 10^{-12}$ | $6.03 \times 10^{-13}$ |
| 298 | $5.27 \times 10^{-11}$ | $3.44 \times 10^{-11}$ | $1.07 \times 10^{-11}$ | $7.08 \times 10^{-12}$ | $5.02 \times 10^{-13}$ |
| 500 | $4.44 \times 10^{-11}$ | $2.85 \times 10^{-11}$ | $9.82 \times 10^{-12}$ | $5.68 \times 10^{-12}$ | $4.80 \times 10^{-13}$ |
| 700 | $4.06 \times 10^{-11}$ | $2.52 \times 10^{-11}$ | $9.85 \times 10^{-12}$ | $5.01 \times 10^{-12}$ | $4.85 \times 10^{-13}$ |
| 1000 | $3.72 \times 10^{-11}$ | $2.20 \times 10^{-11}$ | $1.02 \times 10^{-11}$ | $4.43 \times 10^{-12}$ | $5.83 \times 10^{-13}$ |
| 1500 | $3.43 \times 10^{-11}$ | $1.87 \times 10^{-11}$ | $1.12 \times 10^{-11}$ | $3.78 \times 10^{-12}$ | $6.71 \times 10^{-13}$ |
| 2000 | $3.27 \times 10^{-11}$ | $1.62 \times 10^{-11}$ | $1.22 \times 10^{-11}$ | $3.48 \times 10^{-12}$ | $8.24 \times 10^{-13}$ |
| 2500 | $3.23 \times 10^{-11}$ | $1.48 \times 10^{-11}$ | $1.33 \times 10^{-11}$ | $3.34 \times 10^{-12}$ | $9.26 \times 10^{-13}$ |

Table S4. Product branching ratios (in percent) for the $NH_2$ + O reaction from QCT simulations with the RZPE correction scheme (QCT + RZPE).

| $T$ / K | $f_{\mathrm{HNO+H}}^{\mathrm{RZPE}}$ | $f_{\mathrm{NH+OH}}^{\mathrm{RZPE}}$ | $f_{\mathrm{NO+H_2}}^{\mathrm{RZPE}}$ | $f_{\mathrm{HON+H}}^{\mathrm{RZPE}}$ |
|---|---|---|---|---|
| 200 | 70.02 | 15.00 | 14.34 | 0.59 |
| 298 | 69.40 | 15.82 | 14.07 | 0.60 |
| 500 | 68.07 | 17.66 | 13.38 | 0.70 |
| 700 | 66.25 | 19.87 | 12.87 | 0.87 |
| 1000 | 62.88 | 23.35 | 12.50 | 1.26 |
| 1500 | 57.39 | 29.23 | 11.56 | 1.75 |
| 2000 | 51.98 | 34.36 | 11.07 | 2.46 |
| 2500 | 47.59 | 39.00 | 10.70 | 2.84 |

Table S5. Product branching ratios (in percent) for the $NH_2$ + O reaction from QCT simulations performed without enforcing ZPE conservation (QCT + NZPE).

| $T$ / K | $f_{\mathrm{HNO+H}}^{\mathrm{NZPE}}$ | $f_{\mathrm{NH+OH}}^{\mathrm{NZPE}}$ | $f_{\mathrm{NO+H_2}}^{\mathrm{NZPE}}$ | $f_{\mathrm{HON+H}}^{\mathrm{NZPE}}$ |
|---|---|---|---|---|
| 200 | 65.84 | 19.47 | 13.75 | 1.00 |
| 298 | 65.28 | 20.30 | 13.43 | 0.99 |
| 500 | 64.19 | 22.12 | 12.79 | 1.08 |
| 700 | 62.07 | 24.26 | 12.34 | 1.19 |
| 1000 | 59.14 | 27.42 | 11.91 | 1.57 |
| 1500 | 54.52 | 32.65 | 11.02 | 1.96 |
| 2000 | 49.54 | 37.31 | 10.64 | 2.52 |
| 2500 | 45.82 | 41.18 | 10.34 | 2.87 |

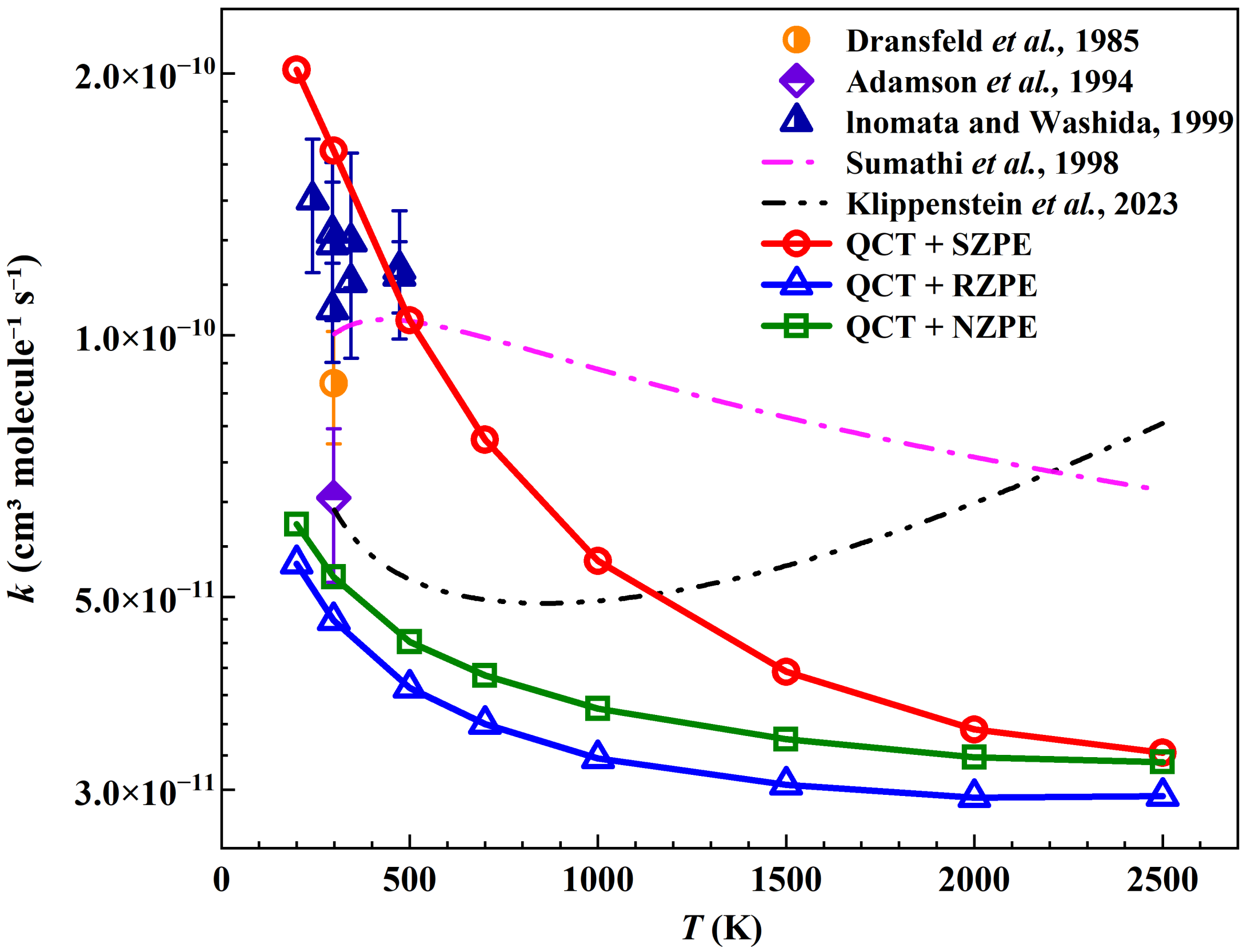

Figure S1. Comparison of total thermal rate coefficients for the $NH_2$ + O reaction from QCT simulations under different ZPE treatments: employing the SZPE correction (QCT + SZPE), employing the RZPE correction (QCT + RZPE), and without enforcing ZPE conservation (QCT + NZPE). Available experimental[1-3] and theoretical[4, 5] data from literatures are also plotted for comparison.

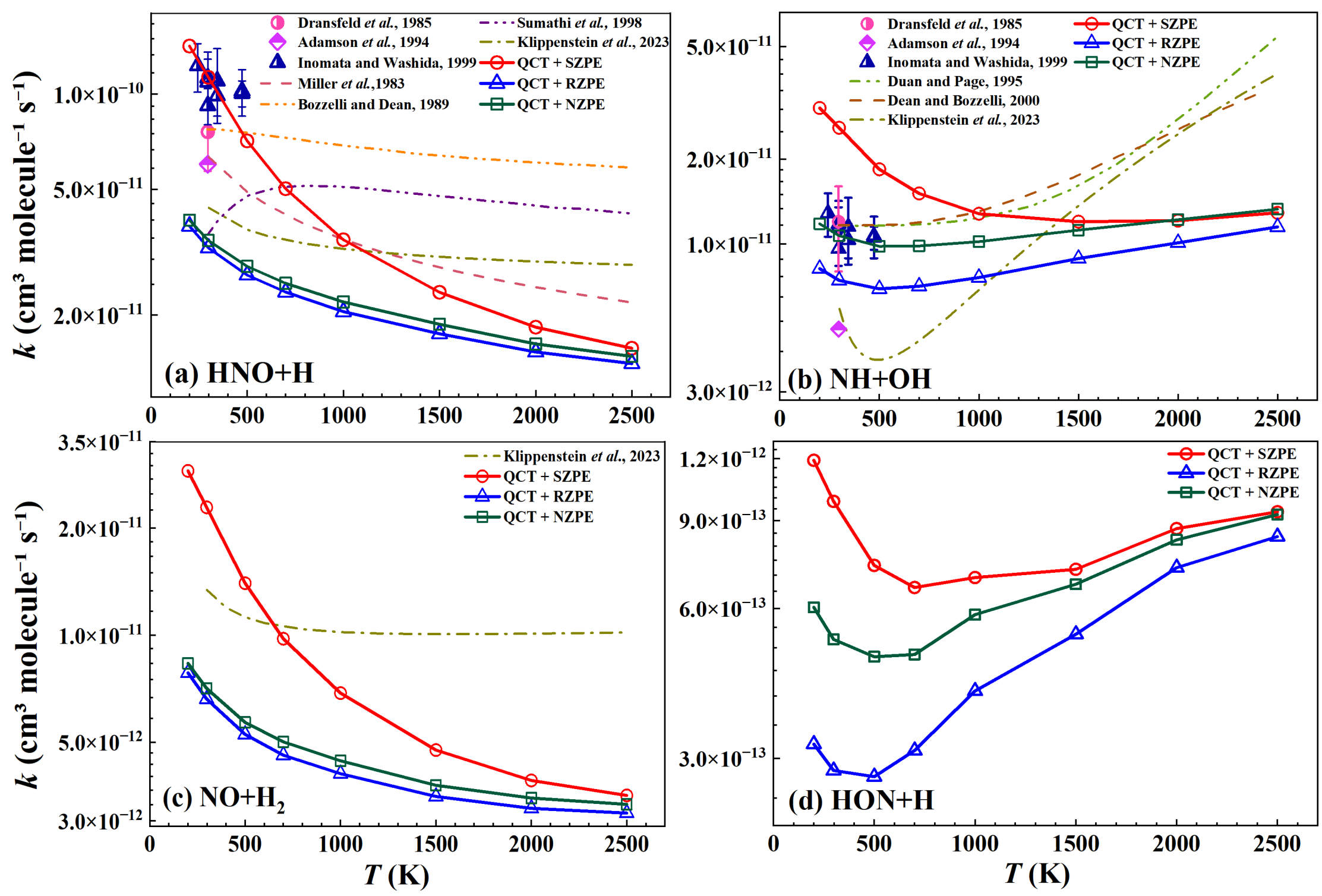

Figure S2. Channel-specific thermal rate coefficients for the (a) HNO + H, (b) NH + OH, (c) NO + $H_2$, and (d) HON + H reaction channels under the same ZPE treatment schemes as in Figure S1. Available experimental[1-3] and theoretical[4-9] data from literatures are included for comparison.

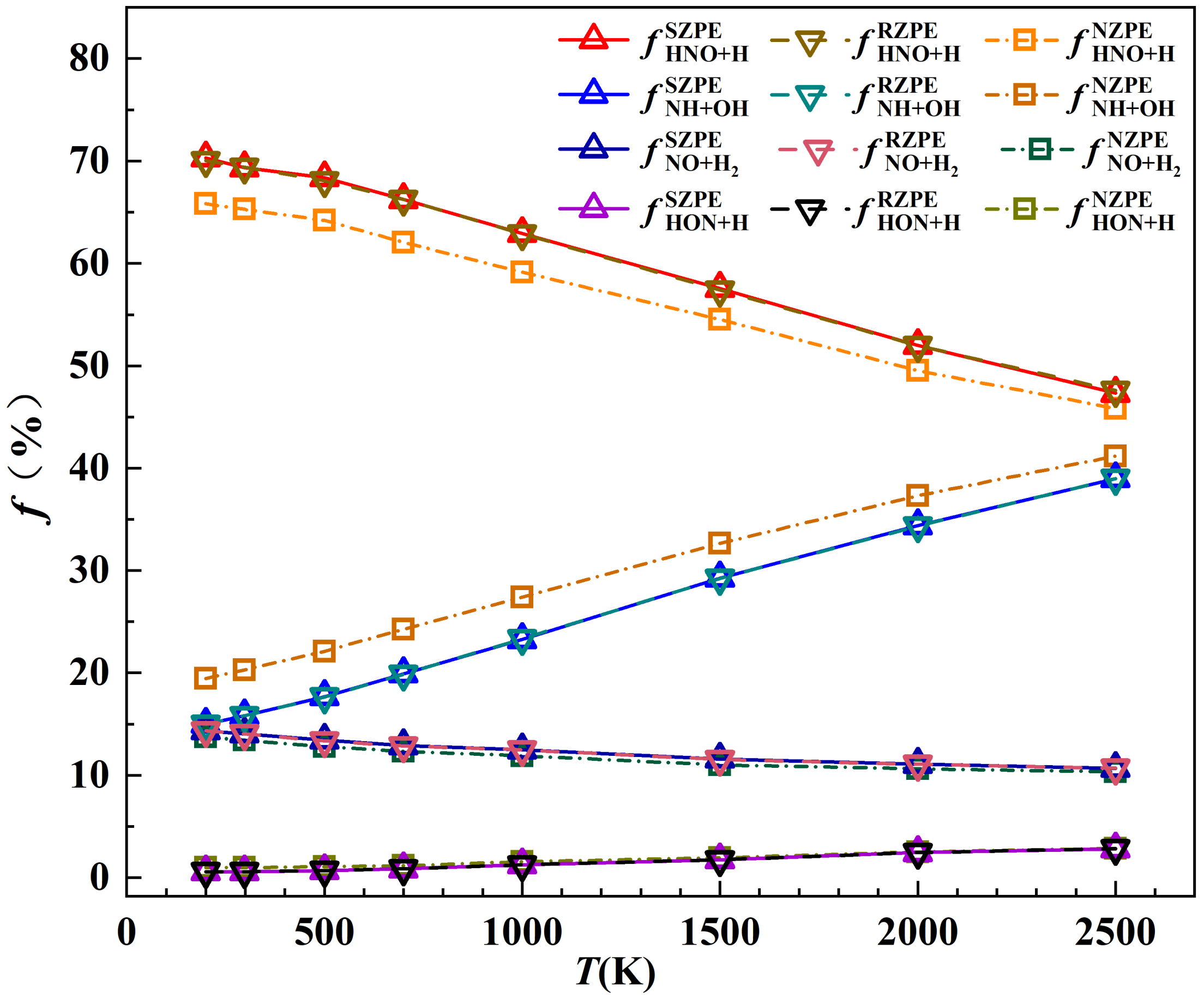

Figure S3. Comparison of product branching ratios for the $NH_2$ + O reaction from QCT simulations under the same ZPE treatment schemes as in Figure S1.

## S2. Cartesian Coordinates

The Cartesian coordinates (in Å) for all identified stationary points (as illustrated in Figure 3), optimized at the MRCI-F12/cc-pVTZ-F12 level of theory.

| | | | |
|---|---|---|---|
| $NH_2$ | | | |
| N | -4.78701814 | -0.36839264 | 0.00000000 |
| H | -5.16074093 | 0.58530213 | 0.00000000 |
| H | -3.77401701 | -0.21663277 | 0.00000000 |
| HNO | | | |
| N | -4.26051495 | 0.32605831 | 0.29935634 |
| O | -4.26331698 | 0.34066032 | -0.91150561 |
| H | -4.26405625 | -0.66677848 | 0.61214154 |
| HON | | | |
| N | -3.88053445 | -0.18527817 | -0.70537452 |
| O | -3.98463772 | 0.13114130 | 0.58060353 |
| H | -4.92040517 | 0.05417191 | 0.81413136 |
| NH | | | |
| N | 5.54459067 | 0.00000000 | -0.51812151 |
| H | 5.54454062 | 0.00000000 | 0.51813508 |
| OH | | | |
| O | -5.54474592 | 0.00000000 | -0.48443949 |
| H | -5.54438537 | 0.00000000 | 0.48442593 |
| NO | | | |
| N | -5.49934339 | 0.00000000 | 0.57590081 |
| O | -5.49940220 | 0.00000000 | -0.57590517 |
| $H_2$ | | | |
| H | 5.49938979 | 0.00000000 | 0.37149279 |
| H | 5.49935581 | 0.00000000 | -0.37148843 |
| $H_2NO$ | | | |
| N | 0.06165435 | -0.01210978 | 0.09308755 |
| O | -0.23838460 | 0.99119625 | -0.63280271 |
| H | 0.80262691 | -0.61733984 | -0.20911219 |
| H | -0.62589666 | -0.36174663 | 0.74882735 |
| *Trans*-HNOH | | | |
| N | -0.62171442 | -0.02000481 | -0.30703568 |
| O | 0.59999112 | -0.00637085 | 0.31428596 |
| H | -0.56017598 | 0.82515978 | -0.86702910 |
| H | 0.58189928 | -0.79878412 | 0.85977883 |
| *cis*-HNOH | | | |
| N | -0.57175288 | 0.30336622 | 0.53792740 |
| O | 0.63572560 | -0.32920086 | 0.40407740 |
| H | -0.90249805 | 0.46866202 | -0.41085385 |
| H | 0.83852533 | -0.44282738 | -0.53115094 |

| TS1 | | | |
|---|---|---|---|
| N | -0.15113855 | -0.46307205 | 0.13823393 |
| O | -0.17790305 | 0.69135175 | -0.67653398 |
| H | -0.41935275 | -0.15910745 | 1.07364793 |
| H | 0.74839435 | -0.06917225 | -0.53534788 |
| TS2 | | | |
| N | -0.58953260 | 0.32466957 | 0.42856386 |
| O | 0.65311731 | -0.32635515 | 0.28683165 |
| H | -1.17789697 | -0.14654818 | -0.28212398 |
| H | 1.11431226 | 0.14823376 | -0.43327154 |
| TS3 | | | |
| N | 0.55128973 | -0.12063340 | -0.02682183 |
| O | 0.57478373 | 0.92675020 | 0.63081838 |
| H | -0.52611408 | -0.14164950 | -0.71964223 |
| H | -0.59995938 | -0.66446730 | 0.11564568 |
| TS4 | | | |
| N | -0.56668000 | -0.51136330 | -0.01793954 |
| O | 0.61533941 | -0.21242195 | 0.18324026 |
| H | -0.64112400 | -0.66724055 | -1.03789014 |
| H | 0.59246459 | 1.39102581 | 0.87258942 |
| TS5 | | | |
| N | 0.04006253 | -0.29731745 | -0.26690900 |
| O | -0.44645658 | -1.36829555 | 0.00117610 |
| H | 0.96487723 | -0.18671245 | 0.24102160 |
| H | -0.55848318 | 1.85232545 | 0.02471130 |

**References**

[1] P. Dransfeld, W. Hack, H. Kurzke, F. Temps, and H. G. Wagner, Direct studies of elementary reactions of $NH_2$-radicals in the gas phase, Chem. Phys. Lett. **20**, 655–663 (1985).

[2] J. D. Adamson, S. K. Farhat, C. L. Morter, G. P. Glass, R. F. Curl, and L. F. Phillips, The Reaction of $NH_2$ with O, J. Phys. Chem. **98**, 5665–5669 (1994).

[3] S. Inomata, and N. Washida, Rate Constants for the Reactions of $NH_2$ and HNO with Atomic Oxygen at Temperatures between 242 and 473 K, J. Phys. Chem. A **103**, 5023–5031 (1999).

[4] S. J. Klippenstein, C. R. Mulvihill, and P. Glarborg, Theoretical Kinetics Predictions for Reactions on the NH2O Potential Energy Surface, J. Phys. Chem. A **127**, 8650–8662 (2023).

[5] R. Sumathi, D. Sengupta, and M. T. Nguyen, Theoretical Study of the $H_2$ + NO and Related Reactions of [$H_2$NO] Isomers, J. Phys. Chem. A **102**, 3175–3183 (1998).

[6] J. A. Miller, M. D. Smooke, R. M. Green, and R. J. Kee, Kinetic Modeling of the Oxidation of Ammonia in Flames, Combust. Sci. Technol. **34**, 149–176 (1983).

[7] J. Bozzelli, and A. Dean, Energized complex quantum Rice-Ramsperger-Kassel analysis of reactions of $NH_2$, $O_2$, and O atoms, J. Phys. Chem. **93**, 1058–1065 (1989).

[8] X. Duan, and M. Page, *Ab initio* variational transition state theory calculations for the O+$NH_2$ hydrogen abstraction reaction on the $^4A'$ and $^4A''$ potential energy surfaces, J. Chem. Phys. **102**, 6121–6127 (1995).

[9] A. M. Dean, and J. W. Bozzelli, in *Gas-phase combustion chemistry* (Springer, 2000), pp. 125–341.